\pdfoutput=1
\documentclass[journal,twoside,web]{ieeecolor}
\usepackage{generic}
\usepackage{cite}
\usepackage{amsmath,amssymb,amsfonts}

\usepackage{algpseudocode}

\usepackage{threeparttable}  

\usepackage{graphicx}
\usepackage{textcomp}
\usepackage{multirow}
\usepackage{ragged2e}
\usepackage{bbm}
\usepackage{booktabs}
\usepackage{makecell}
\usepackage{array}
\usepackage{xcolor}
\usepackage{tikz}
\usepackage{xcolor}
\usepackage{hyperref}
\hypersetup{
    colorlinks=true,
    citecolor=green,
    linkcolor=magenta,
    filecolor=blue,      
    urlcolor=blue,
}

\def\BibTeX{{\rm B\kern-.05em{\sc i\kern-.025em b}\kern-.08em
    T\kern-.1667em\lower.7ex\hbox{E}\kern-.125emX}}

\begin{document}
\title{Phi-SegNet: Phase-Integrated Supervision for Medical Image Segmentation}
\author{Shams~Nafisa~Ali,~\IEEEmembership{Student~Member, IEEE} and Taufiq~Hasan,~\IEEEmembership{Senior~Member, IEEE}\\
\thanks{S.~N.~Ali and T. Hasan are with the mHealth Lab, Department of Biomedical Engineering, Bangladesh University of Engineering and Technology, Bangladesh, e-mail: \{nafisa, taufiq\}@bme.buet.ac.bd}
\thanks{S.~N.~Ali is also affiliated with the Department of Electrical and Computer Engineering, Johns Hopkins University, Baltimore, MD, USA.}%
\thanks{T. Hasan has a secondary affiliation with the Center for Bioengineering Innovation and Design, Department of Biomedical Engineering, Johns Hopkins University, Baltimore, MD, USA.}}

\maketitle

\begin{abstract}
Deep learning has substantially advanced medical image segmentation, yet achieving robust generalization across diverse imaging modalities and anatomical structures remains a major challenge. A key contributor to this limitation lies in how existing architectures—ranging from Convolutional Neural Networks (CNNs) to Transformers and their hybrids—primarily encode spatial information while overlooking frequency-domain representations that capture rich structural and textural cues. Although few recent studies have begun exploring spectral information at the feature level, supervision-level integration of frequency cues—crucial for fine-grained object localization—remains largely untapped. To this end, we propose \textbf{\textit{Phi-SegNet}}, a CNN-based architecture that incorporates phase-aware information at both architectural and optimization levels. The network integrates \textit{Bi-Feature Mask Former (BFMF)} modules that blend neighboring encoder features to reduce semantic gaps, and \textit{Reverse Fourier Attention (R$\mathcal{F}$A)} blocks that refine decoder outputs using phase-regularized features. A dedicated phase-aware loss aligns these features with structural priors, forming a closed feedback loop that emphasizes boundary precision. Evaluated on five public datasets spanning ultrasound, X-ray, histopathology, MRI, and colonoscopy, Phi-SegNet consistently achieved state-of-the-art performance, with an average relative improvement of $1.54\,\pm 1.26\%$ in IoU and $0.98\,\pm 0.71\%$ in F1-score over the next best-performing model. Additionally, in cross-dataset generalization scenarios involving unseen datasets from the known domain, Phi-SegNet exhibits robust and superior performance—highlighting its adaptability and modality-agnostic design. These findings demonstrate the potential of leveraging spectral priors in both feature representation and supervision, paving the way for generalized segmentation frameworks that excel in fine-grained object localization. The source code is available on \href{https://github.com/ShamsNafisaAli/Phi-SegNet}{GitHub}.
\end{abstract}

\begin{IEEEkeywords}
Medical image segmentation, deep learning, frequency domain, phase, reverse attention.
\end{IEEEkeywords}

\section{Introduction}
\label{sec:introduction}
\IEEEPARstart{M}{edical} image segmentation (MIS)—the task of identifying and delineating regions of interest such as organs, tumors, or lesions from medical images—plays a pivotal role in accurate diagnosis, treatment planning, and disease monitoring. However, manual segmentation demands substantial time and expert anatomical knowledge, resulting in high inter-observer variability and limited reproducibility. Consequently, there is an increasing demand for automated and reliable segmentation tools that can improve both the precision and throughput of clinical workflows.

Over the last decade, deep learning (DL) approaches, particularly those based on convolutional neural networks (CNNs), have significantly advanced MIS by learning rich feature hierarchies from data, outperforming traditional, hand-engineered methods. Among CNN-based architectures, U-Net~\cite{unet} introduced a pivotal encoder-decoder framework with symmetric skip connections that fuse low-level spatial detail with high-level semantic context. However, the simplicity of U-Net limits its effectiveness when dealing with small or irregular structures, closely situated targets, fuzzy boundaries, and visually ambiguous regions. To address these challenges, numerous variants have been developed that incorporate multi-scale feature aggregation~\cite{cpfnet, msrfnet, m2snet}, redesigned skip pathways~\cite{unet++}, and advanced attention mechanisms~\cite{att1, eanet, multi1, aaunet}. Building on these advances, CE-Net~\cite{cenet} and CPFNet~\cite{cpfnet} expanded the receptive field of the encoder through dense atrous convolution and multiscale context pooling, enabling more effective extraction of local and global semantic characteristics. Furthermore, models such as PraNet~\cite{pranet} and PRCNet~\cite{prcnet} employ reverse attention to iteratively refine coarse predictions and enhance the delineation of ambiguous boundaries. Complementing these designs, dual-branch frameworks such as Twin-SegNet~\cite{twinsegnet} improve segmentation consistency by jointly learning from both foreground and background perspectives.

Despite their success, CNNs are inherently limited by their localized receptive fields, which restrict their capacity to model long-range dependencies, crucial for understanding global anatomical context. This limitation has motivated the integration of transformer architectures in MIS. Early models like TransUNet~\cite{transunet} combine convolutional encoders with vision transformers to capture global relationships among features, while hybrid designs such as Transfuse~\cite{transfuse}, TransAttUNet~\cite{transattunet}, and H2Former~\cite{h2former} introduce hierarchical or scale-aware attention to enhance semantic consistency. These architectures have shown improved performance, particularly in complex anatomical scenes, but often demand large annotated datasets and rely heavily on spatial-domain cues.

As spatial-domain methods continue to dominate the field, frequency-domain modeling has recently emerged as a valuable complement by offering global signal representations through amplitude and phase decomposition. This perspective is clinically relevant in segmentation scenarios where local spatial appearance alone is ambiguous. Examples include volumetric structures with diffuse margins, such as prostate or abdominal organs in MRI/CT; thin, tubular or elongated structures, such as retinal vessels, angiograms, and nailfold capillaries; irregular lesions, such as breast tumors in ultrasound or skin lesions; and densely packed fine structures, such as cells and glands in histopathology or teeth in dental radiographs. In these cases, target regions may exhibit weak boundary contrast, scale variation, complex topology, or intensity similarity with surrounding tissues~\cite{han_fass, fdenet}. Consequently, purely spatial-domain models may produce incomplete foreground masks, boundary discontinuities, topology errors, or leakage into anatomically similar background regions. Techniques based on Fourier, cosine, or wavelet transforms can help mitigate these limitations by emphasizing structural regularities, frequency-band-specific details, aliasing-reduced representations, foreground-relevant components, and boundary/detail cues that are not always sufficiently expressed in local intensity patterns.

Recent frequency-spatial and frequency-aware frameworks show that spectral information can support segmentation through frequency-domain feature extraction, frequency-band selection, wavelet-based downsampling, aliasing reduction, foreground enhancement, and high-frequency boundary/detail recovery~\cite{fdenet, ew-vit, xing_frequfnet}. MEW-UNet~\cite{mewunet}, FDFUNet~\cite{fdfunet}, and GFUNet~\cite{gfunet} leverage frequency filtering and spectral recalibration to enhance boundary delineation and reduce redundancy. Dual-domain networks such as DBL-Net~\cite{dblnet}, D$^2$LNet~\cite{d2lnet}, and SF-UNet~\cite{sfunet} combine spatial and frequency-domain representations through multi-level fusion or dual-domain attention to better resolve semantic ambiguity. More recent approaches have explored explicit spectral attention and deeper spectral integration. SLf-UNet~\cite{slfunet} introduces spectral recalibration units operating on frequency-transformed features to highlight relevant bands and suppress flat or noisy components, while tKFC-Net~\cite{tkfcnet} applies twin-kernel frequency convolutions at each stage to balance global context with edge precision. Following this direction, BAWGNet~\cite{bawgnet} embeds frequency-aware filtering within global attention pathways, whereas EW-ViT~\cite{ew-vit} incorporates wavelet decomposition into the self-attention block to refine both local detail representation and global contextual modeling in vision transformers.

Nevertheless, most existing frequency-aware methods primarily exploit spectral information at the feature-extraction or feature-enhancement level. Although these strategies can improve representation learning, they do not necessarily impose an explicit structural constraint on decoder outputs, where object boundaries, topology, and final mask geometry are reconstructed. In particular, phase-based guidance during training remains underexplored, and existing architectures rarely establish a feedback mechanism between decoder predictions and spectral properties. This limits their ability to directly regularize boundary continuity, suppress unstable residual responses, and improve decoder-level structural consistency. These observations motivate a complementary and more principled integration of spectral priors within the segmentation pipeline, ideally in a way that informs both attention formation and optimization dynamics while supporting spatial-semantic representation learning across scales.

To address these limitations, we propose \textbf{Phi-SegNet}, a frequency-aware segmentation framework that couples multi-scale spatial-semantic feature learning with spectral reasoning in the attention pathway and supervision process. First, we introduce a \textit{Bi-Feature Mask Former (BFMF)} module that fuses adjacent encoder features to reduce the semantic gap between low-level spatial detail and high-level contextual information, providing the decoder with coherent multi-scale representations. Second, we propose a \textit{Reverse Fourier Attention (R$\mathcal{F}$A)} mechanism that refines decoder features by filtering reverse attention maps in the frequency domain, selectively suppressing unstable residual responses while preserving spatially coherent regions that require refinement. Finally, we incorporate a novel \textit{phase-integrated supervision loss}, which enforces Fourier phase alignment between phase-conditioned decoder features and the ground-truth mask, offering a structural constraint that complements the spatial pixel-wise loss. These components collectively enable Phi-SegNet to exploit both spatial and spectral cues for robust, boundary-preserving segmentation. Extensive evaluations across five distinct medical imaging modalities, i.e., ultrasound, X-ray, endoscopy, histology, and MRI, demonstrate the superiority of Phi-SegNet over state-of-the-art spatial-domain, frequency-domain, and dual-domain models in terms of overall segmentation performance and boundary fidelity.

\section{Background}\label{sec:background}
Fourier Transform offers a powerful means to analyze spatial patterns in images by decomposing them into frequency components, typically realized through its two-dimensional discrete form. Let $x[m,n]$ be a grayscale image of size $M \times N$. The 2D DFT and its inverse are defined as:
\begin{align}
X[k,l] &= \sum_{m=0}^{M-1} \sum_{n=0}^{N-1} x[m,n] e^{-j \left( \frac{2\pi}{M}km + \frac{2\pi}{N}ln \right)}
\\
x[m,n] &= \frac{1}{MN} \sum_{k=0}^{M-1} \sum_{l=0}^{N-1} X[k,l] e^{j \left( \frac{2\pi}{M}km + \frac{2\pi}{N}ln \right)}
\end{align}
where $k = 0, 1, ..., M-1$ and $l = 0, 1, ..., N-1$.

Each Fourier coefficient \(X[k,l]\) represents the frequency component corresponding to the spatial frequency index \((k,l)\). The \emph{magnitude spectrum} characterizes the energy distribution of these frequency components, capturing texture patterns, intensity modulations, and dominant frequency bands, but it lacks spatial and structural localization. In contrast, the \emph{phase spectrum} encodes the relative alignment of frequencies, preserving spatial structure, contour integrity, and geometric information within the image. Given the complex spectrum $X[k,l]$, the magnitude and phase spectra are respectively defined as:
\begin{align}
|X[k,l]| &= \sqrt{\Re\{X[k,l]\}^2+\Im\{X[k,l]\}^2},\\
\angle X[k,l] &= \arctan\!\left(\frac{\Im\{X[k,l]\}}{\Re\{X[k,l]\}}\right).
\end{align}

Phase governs the spatial alignment of image structures in the frequency domain. 
When an image is translated, its magnitude spectrum remains unchanged, while the phase undergoes a linear shift.  
Let a two-dimensional image \(f(x,y)\) have the Fourier transform \(F(u,v)\):
\begin{equation}
F(u,v)=\iint f(x,y)\,e^{-j2\pi(ux+vy)}\,dxdy.
\end{equation}
A translation by \((x_0,y_0)\) produces a shifted image \(f'(x,y)=f(x-x_0,y-y_0)\) with
\begin{align}
F'(u,v)&=F(u,v)\,e^{-j2\pi(ux_0+vy_0)},\\
|F'(u,v)|&=|F(u,v)|,\\
\angle F'(u,v)&=\angle F(u,v)-2\pi(ux_0+vy_0).
\end{align}
Thus, spatial translation affects only the phase spectrum, confirming that phase information 
preserves positional and structural fidelity, whereas the magnitude spectrum reflects overall frequency content.

Beyond structural representation, frequency-domain modeling also provides an effective mechanism for capturing long-range contextual dependencies~\cite{rao2021global}. Since each Fourier coefficient is computed from all spatial locations of the input image or feature map, spectral representations aggregate information globally rather than relying only on local neighborhoods. Consequently, learning or reweighting frequency components and transforming them back to the spatial domain can be interpreted as a global operation over the entire feature map, as frequency-domain modulation corresponds to spatial-domain filtering under the convolution theorem~\cite{chi2020fast}. This intrinsic global receptive-field property enables spectral cues to encode relationships across distant spatial regions, thereby complementing local convolutional operations in long-range contextual modeling. Therefore, integrating frequency-domain information is particularly beneficial for medical image segmentation, where both local boundary details and long-range anatomical consistency are important for accurate mask prediction.

\section{Methodology}\label{sec:methodology}
\subsection{Baseline Model}
\label{subsec:baseline_model}
An encoder–decoder-based segmentation model with the EfficientNet-B4~\cite{tan2019efficientnet} backbone 
has been selected as the baseline model for this work due to its excellent trade-off between performance and computational cost.

\begin{figure*}
    \centering
    \includegraphics[width=0.89\textwidth]{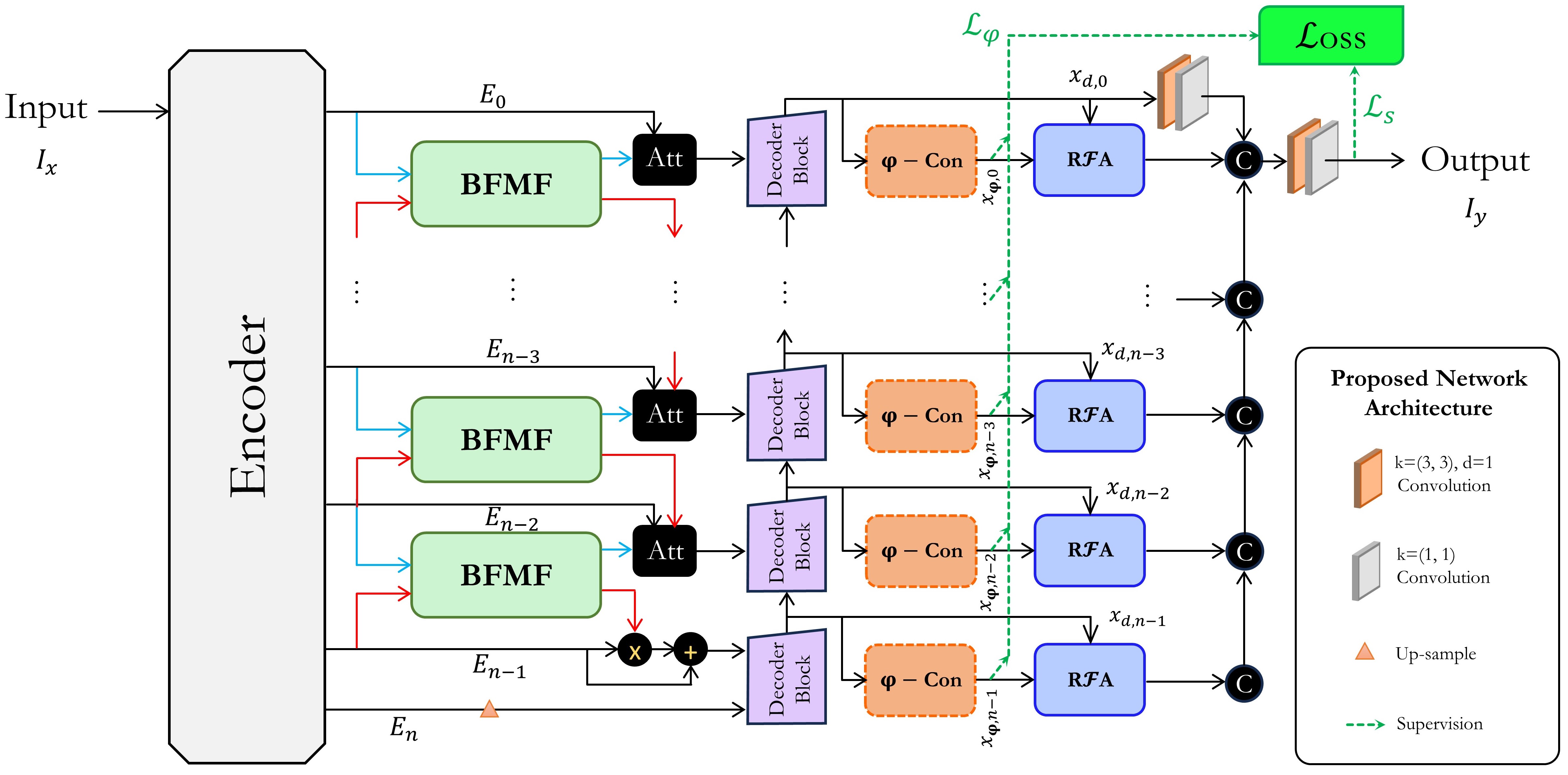}
    \caption{Overview of the proposed Phi-SegNet. The architecture integrates encoder features via bi-feature mask former (BFMF) modules and attention-guided skip connections. Phase supervision is applied on decoder stages followed by reverse Fourier attention (R$\mathcal{F}$A) modules, which use spectral filtering to enhance boundary localization.}
    \label{fig:total_architecture}
\end{figure*}

\subsection{Proposed Phi-SegNet Architecture}
\subsubsection{Encoder Pipeline}
The proposed \textbf{Phi-SegNet} architecture (illustrated in Fig.~\ref{fig:total_architecture}) employs an encoder pipeline that effectively extracts contextual representations relevant to the segmentation task across multiple feature levels. In this work, the input image \(I_x \in \mathbb{R}^{H \times W \times 3}\) is passed through a pre-trained EfficientNet-B4 encoder 
to obtain a set of hierarchical feature maps \(\{E_0, E_1, \dots, E_n\}\), 
where each \(E_i\) represents the feature map at level \(i\) with its corresponding spatial resolution, and \(n\) denotes the total number of spatial hierarchy levels.
\vspace{0.2 cm}

\subsubsection{Bi-Feature Mask Former (BFMF) Module}
\begin{figure}[htbp]
    \centering
    \includegraphics[width=0.9\linewidth]{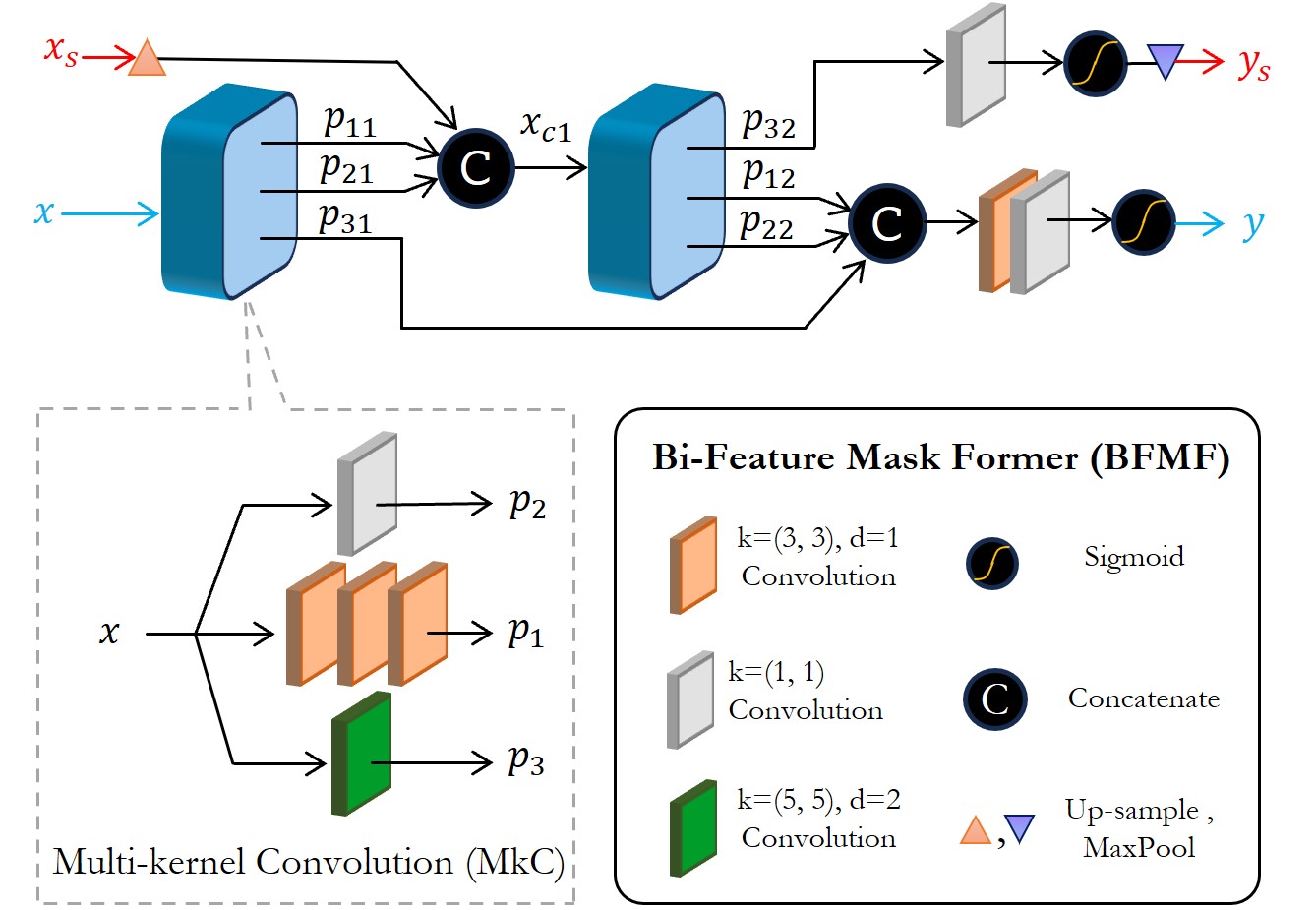}
    \caption{The Bi-Feature Mask Former (BFMF) module extracts multi-scale semantic and spatial features through multi-kernel convolutions (1×1, 3×3, 5×5) and sequential aggregation.}
    \label{fig:bfmf}
\end{figure}

The Bi-Feature Mask Former (BFMF) module integrates feature representations from two adjacent encoder levels, \(E_i\) and \(E_{i+1}\), to enhance the fusion of semantic and spatial context. 
As illustrated in Fig.~\ref{fig:total_architecture}, the blue arrows feeding into the BFMF denote higher-resolution encoder features \(E_i\), whereas the red arrows correspond to the adjacent lower-resolution encoder features \(E_{i+1}\). In Fig.~\ref{fig:bfmf}, we respectively denote these as \(x \) and \(x_s\) for notational simplicity. 

Given \(x \in \mathbb{R}^{H \times W \times C}\) and \(x_s \in \mathbb{R}^{\tfrac{H}{2} \times \tfrac{W}{2} \times C_s}\), 
the module produces two corresponding mask features, \(y\) and \(y_s\), at the same spatial scales. Multi-kernel convolutions (MkC) with varying kernel sizes (\(k=\{1,3,5\}\)) and dilation rates (\(d=\{1,2\}\)) are employed to capture multi-scale receptive fields, as illustrated in Fig.~\ref{fig:bfmf}. Features from higher dilation rates are processed less to better conserve their spatial dependencies. The operations are formulated as:
\begin{align}
p_{11}, p_{21}, p_{31} &= \text{MkC}(x), \\
x_{c1} &= p_{11} \oplus p_{21} \oplus \text{Up}(x_s), \\
p_{12}, p_{22}, p_{32} &= \text{MkC}(x_{c1}), \\
y &= \sigma\!\left((p_{12} \oplus p_{22} \oplus p_{31}) * W_3^1 * W_1^1\right), \\
y_s &= \text{Max}\!\left(\sigma(p_{32}) * W_1^1\right),
\end{align}
where \(W_k^d\) denotes a learnable weight tensor (k = kernel size, d = dilation rate), \(\oplus\) represents feature concatenation, 
\(\sigma\) is the sigmoid activation function, and
\( \ast \) indicates convolution. 

The resulting mask features \( y \in \mathbb{R}^{H \times W \times C}\) and \( y_s \in \mathbb{R}^{\tfrac{H}{2} \times \tfrac{W}{2} \times C_s}\) are then passed through attention modules (Att) to generate refined masks and further refine the encoded features \(E_i \;\forall i \in [0, n-2]\).

To capture both fine-grained information and global context, the BFMF module leverages cascaded and dilated convolutions to expand the effective receptive field (RF) without additional parameters or resolution loss. Stacking three \(3 \times 3\) convolutional layers (with stride 1 and padding) results in an effective RF of \(7 \times 7\), enabling hierarchical aggregation of local features. In contrast, a single \(5 \times 5\) convolution with \(d = 2\) achieves a receptive field of \(9 \times 9\), allowing it to directly access broader spatial context with fewer operations. By combining both strategies, BFMF benefits from a dual perspective: cascaded layers extract fine-grained details progressively, while dilated convolutions enhance global awareness through sparse, long-range connections. This design empowers the module to generate masks that encode both localized boundaries and long-range dependencies.
\subsubsection{Attention Module}
\begin{figure}[b]
    \centering
    \includegraphics[width=\columnwidth]{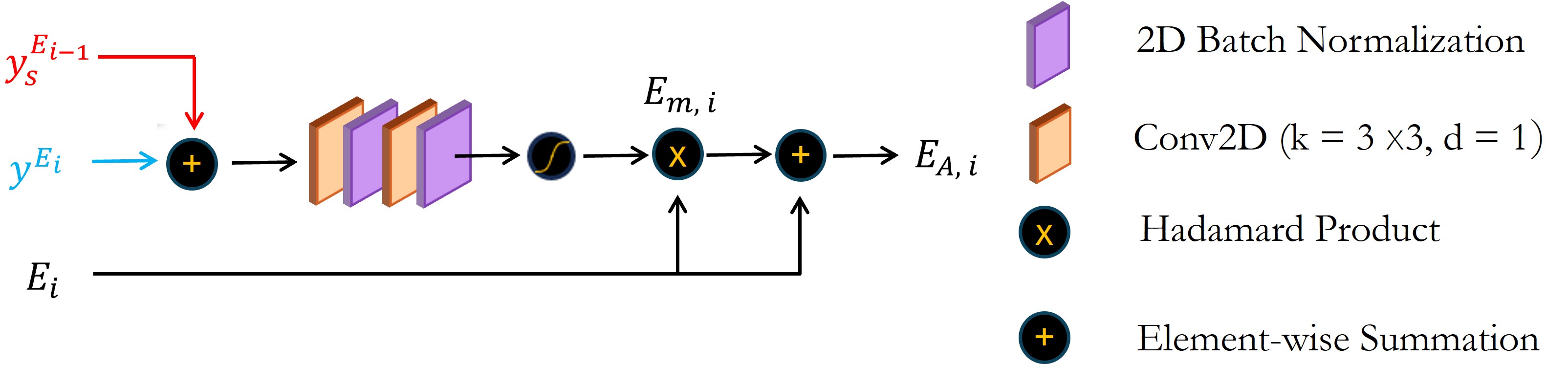}
    \caption{Attention-guided fusion strategy. At each encoder stage \(E_i\), the attention mask \(y^{E_i}\) is combined with the max-pooled mask from the preceding stage \(y^{E_{i-1}}_s\), and passed through a convolutional block followed by a sigmoid activation to generate a refined attention map. This map modulates the encoder features via element-wise multiplication and is subsequently added to the decoder pathway. 
    }
    \label{fig:att}
\end{figure}

The attention modules (see Fig.~\ref{fig:att}) concatenate adjacent level features $y_s^{E_{i-1}}$ and $y^{E_i}$ to extract spatial information which are optimum for refining the encoded quantity \(E_i\). The generation of the refinement mask $E_{m,i}$ and the attention process can be shown as follows:

\begin{equation}
   E_{m,i} = \sigma\!\left(BN\!\left(BN\!\left((y_s^{E_{i-1}} \oplus y^{E_i}) * W_3^1\right) * W_3^1\right)\right)
\end{equation}
\begin{equation}
   E_{A,i} = (E_i \otimes E_{m,i}) + E_{i}
\end{equation}
where $BN(.)$ indicates 2D-Batch Normalization, \( \otimes \) and \( + \) denote the Hadamard product and element-wise summation, respectively. 

\subsubsection{Decoder Module}
The decoder blocks take the output from the attention module, \( E_{A,i} \), and concatenate it with the previous decoder output \( x_{d,i+1} \). The concatenated features are then upsampled to match the resolution of the feature map at the current level, followed by a DoubleConv block to reconstruct spatial details and refine boundaries. Each DoubleConv block consists of two sequential convolutional layers, each with a kernel size of \(3 \times 3\), followed by Batch Normalization and a LeakyReLU activation function. This composition facilitates gradient flow and enhances the non-linearity required for accurate reconstruction. 
\[
x_{d,i} = \text{Up}\!\left(\text{DoubleConv}\!\left(E_{A,i} \oplus x_{d,i+1}\right)\right)
\]

The final output of each decoder stage is progressively passed through to gradually decode the segmentation mask.

\subsubsection{$\varphi$-Conditioner}
The phase (\(\varphi\))-conditioner module, shown in Fig.~\ref{fig:total_architecture}, is designed to provide boundary-oriented supervision for the regions of interest. 
Each module consists of a single convolutional layer that projects multi-channel feature tensors into a single-channel representation, followed by a sigmoid activation. This yields feature-level boundary masks \(x_{\varphi,i} \in [0,1]^{H \times W \times 1}\) at each decoder level.
\vspace{0.2 cm}

\begin{figure}
    \centering
    \includegraphics[width=0.99\linewidth]{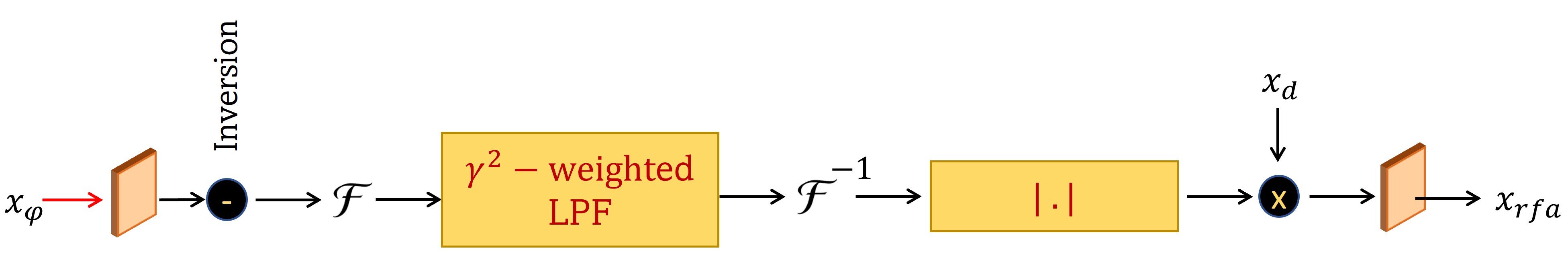}
    \caption{The Reverse Fourier Attention (R$\mathcal{F}$A) module applies a $\gamma^2$-weighted low-pass filter in the frequency domain to enforce phase-conditioned boundary enhancement. Reverse attention focuses on what the model misses and helps it refine low-confidence regions like boundaries or tiny lesions. Early decoder features are coarse, applying LPF in Fourier space suppresses high-frequency noise, allowing attention to act on structurally dominant, globally coherent features.}
    \label{fig:rfa}
\end{figure}

\subsubsection{Reverse Fourier Attention (R$\mathcal{F}$A)} \label{rfa_module}
To supervise the spatial quantities of the segmentation task, we utilize boundary information from the phase-conditioned masks \(x_{\varphi}\) and derive their reverse counterparts to emphasize the complementary foreground and background regions delineated by the boundary. 
The reverse mask is low-pass filtered in the frequency domain by retaining only a centered \(\gamma \times \gamma\) region of low-frequency coefficients and setting the remaining components to zero. 
Such frequency-domain filtering is particularly effective in the early decoder stages, where feature maps are spatially coarse and prone to high-frequency noise.  
The operations within the R$\mathcal{F}$A module are defined as:
\begin{equation}
\hat{x}_{\varphi,i} = 
\left| 
\mathcal{F}^{-1} 
\left( 
\gamma^2 \otimes \mathcal{F}(1.0 - x_{\varphi,i}) 
\right) 
\right|
\end{equation}
where \(\mathcal{F}\) and \(\mathcal{F}^{-1}\) denote the Fourier and inverse Fourier Transforms, respectively; 
\(|\cdot|\) indicates the absolute value; 
and \(\gamma\) is the scalar quantity controlling the cutoff of the low-pass filter (LPF). The resulting feature masks \(\hat{x}_{\varphi,i}\) are subsequently combined with the decoder features \(x_{d,i}\):
\begin{equation}
x_{rfa,i} = (x_{d,i} \otimes \hat{x}_{\varphi,i}) * W_3^1,
\end{equation}
The outputs from all R$\mathcal{F}$A modules are aggregated in the final layer to generate the prediction mask 
\(I_y \in \mathbb{R}^{H \times W \times 1}\).

\medskip

One of the core motivations behind the design of the R$\mathcal{F}$A module is to facilitate complementary learning that explicitly refines the decoder's prediction by suppressing non-salient structures—particularly false positives in background regions. In conventional decoders, repeated fusion and upsampling operations often introduce spatial ambiguity. High-level features are spatially coarse, and their upsampled representations tend to leak into background areas.  Although convolution and interpolation inherently behave as low-pass filters, their effects are implicit and uncontrolled, often failing to selectively suppress uncertain responses.

In contrast, the R$\mathcal{F}$A module applies a deliberate, \(\gamma^2\)-weighted LPF in the frequency domain on the residual between input and decoder prediction. 
Importantly, this filtering is not applied to the final segmentation mask. Rather, it is applied to the reverse attention residual map derived from the phase-conditioned decoder output. Since early decoder features are spatially coarse and can contain fragmented activations, checkerboard-like artifacts, and isolated false-positive responses, the LPF suppresses unstable high-frequency residuals while retaining the dominant spatially coherent regions requiring refinement. Boundary localization is therefore not discarded, as contour fidelity is separately enforced through the phase-aware supervision loss in the Fourier phase domain.
This targeted filtering achieves two complementary refinements: 
(1) enforcing structural consistency by penalizing frequency-domain residuals between predicted and true boundaries, and 
(2) suppressing spurious activations in non-ROI regions while maintaining boundary sharpness. As a result, R$\mathcal{F}$A guides the network to learn not only \emph{what to activate} but also \emph{what to suppress}, significantly reducing false positives and enhancing structural fidelity.

\subsection{Phase-Integrated Loss Function}\label{loss}
Phase information encapsulates the spatial localization and structural geometry of objects, making it crucial for delineating precise boundaries in segmentation tasks. 
To leverage this property, we impose a phase-alignment constraint between the $\varphi$-conditioned features and the corresponding ground-truth mask.  Specifically, we enforce their 2D phase spectra to be consistent in the $L_2$ sense, thereby encouraging the predicted features to preserve boundary coherence in the frequency domain.  However, the phase component of the Fourier transform lies within the range \([0, 2\pi]\), introducing discontinuities and periodicity in the loss landscape.  To mitigate this, the phase terms are unwrapped along both spatial axes prior to loss computation, ensuring that the supervisory signal increases smoothly as the prediction deviates spatially.  The resulting phase-integrated loss is formulated as:
\[
\varphi_{K,i} = \mathbf{u} \left[ \measuredangle x_{\varphi,i}; k \right]; \hspace{3mm}
\varphi_{L,i} = \mathbf{u} \left[ \measuredangle x_{\varphi,i}; l \right]
\]
\[
\mathcal{L}_{\varphi} = \sum_{i=0}^{n-1} \Big\| \varphi_{K,i} - \mathbf{u} \left[ \measuredangle I_y; k \right] \Big\|_2 + \Big\| \varphi_{L,i} - \mathbf{u} \left[ \measuredangle I_y; l \right] \Big\|_2
\]
\noindent where\[
\mathbf{u} \left[P; A\right] = \text{unwrap } P \text{ along A-axis};  
\hspace{2mm}\left\| \cdot \right\|_2 = L_2\text{-Norm}
\]

It should be noted that the phase unwrapping operation is used only during training for computing the phase-supervision loss and is not involved during inference. Therefore, it does not introduce any additional computational overhead at test time. The purpose of unwrapping is to remove artificial \(2\pi\) discontinuities in the wrapped phase map before computing the $L_2$ phase loss. Without this step, two spatially adjacent phase values near the wrapping boundary may produce a large numerical error despite representing nearby structures, resulting in unstable or misleading gradients. By unwrapping the phase along both spatial axes, the phase loss provides a smoother structural alignment signal for backpropagation.

To supervise the spatial accuracy of the predicted mask, we adopt a conventional spatial loss \((\mathcal{L}_s)\) based on the Intersection-over-Union (IoU) criterion:
\begin{equation}
\mathcal{L}_{s} = 
\frac{\sum_{h,w} I_y(h,w) \cdot \hat{I}_y(h,w)}
{\sum_{h,w} I_y(h,w) + \hat{I}_y(h,w) - I_y(h,w) \cdot \hat{I}_y(h,w)}
\end{equation}

The overall objective combines the phase-integrated supervision loss \((\mathcal{L}_\varphi)\) and the spatial IoU loss \((\mathcal{L}_s)\), forming the total loss:
\begin{equation}
\mathcal{L}_{\text{total}} = \alpha \mathcal{L}_\varphi + \beta \mathcal{L}_s
\end{equation}
where \(\alpha\) and \(\beta\) control the relative influence of phase- and spatial-domain supervision.  The configuration \((\alpha = 0.01,\, \beta = 1.00)\) achieved the best trade-off between boundary fidelity and overall segmentation accuracy, producing sharper object contours and fewer background activations across modalities, and was therefore adopted for all subsequent experiments.

\section{Experiments and Results} \label{sec:experimental_result}
\subsection{Datasets}
We conduct experiments on five benchmark tasks involving segmentation of breast lesions (BUSI, Mendeley, UDIAT), teeth (TDD), polyps (Kvasir-SEG, CVC-ColonDB), colorectal glands (GLaS), and prostate (PROMISE-12) from medical images, spanning five widely used imaging modalities such as ultrasound (US), X-ray, colonoscopy, histopathology, and magnetic resonance imaging (MRI). These datasets collectively represent diverse clinical scenarios and modality-specific challenges, including intra-patient variability and heterogeneous acquisition conditions in ultrasound, complex anatomical and textural patterns in dental radiographs, high illumination and morphological variability in endoscopy, stain and tissue heterogeneity in histopathology, and inter-scanner domain shifts in MRI. To evaluate cross-domain robustness, the Mendeley, UDIAT, and CVC-ColonDB datasets are excluded from training and used solely for external validation. For datasets with official splits provided by their organizers (e.g., GLaS and PROMISE-12), those partitions are followed. Otherwise, images are randomly divided into training, validation, and test subsets in an 8:1:1 ratio. All experiments follow identical pre-processing, training, and evaluation protocols. A summary of the datasets is presented in Table~\ref{tab:dataset_summary}. 

\subsection{Experimental Setup}
\subsubsection{Implementation Details}
All experiments have been implemented using PyTorch~2.3.0 with CUDA~12.1 and conducted on an Ubuntu~20.04 workstation equipped with four NVIDIA GeForce RTX~2080 Ti GPUs (11\,GB each), an Intel\textsuperscript{\textregistered} Core\textsuperscript{TM} i9-7920X CPU @ 2.90\,GHz, and 125\,GB of RAM. 

The input images are resized to $256 \times 256$ pixels. Data augmentation strategies included random horizontal and vertical flips, as well as affine transformations with translation. A multi-scale training strategy \{0.5, 1.25\}  is also employed for effectively enforcing scale invariance~\cite{pranet}. The model is trained for 150 epochs using the Adam optimizer with an initial learning rate of $1 \times 10^{-5}$ with a batch size of 4. A cosine annealing learning rate scheduler is applied with $T_{\text{max}}=25$ and $\eta_{\text{min}}=1 \times 10^{-7}$. Model weights corresponding to the lowest validation loss have been retained.

\begin{table}[b]
\centering
\begin{threeparttable}
\caption{Summary of the Datasets Used}
\label{tab:dataset_summary}
\renewcommand{\arraystretch}{1.05}
\setlength{\tabcolsep}{5pt}
\begin{tabular}{lccccc}
\toprule
\textbf{Dataset} & \textbf{Modality} & \textbf{Total} & \textbf{Train} & \textbf{Val} & \textbf{Test} \\
\midrule
\textsuperscript{$\dagger$}BUSI~\cite{busi}            & Breast US & 647  & 517 & 65  & 65  \\
\textsuperscript{$\ddagger$}TDD~\cite{tufts}  & Dental X-ray     & 1000 & 800 & 100 & 100 \\
Kvasir-SEG~\cite{kvasir}                             & Colonoscopy    & 1000 & 800 & 100 & 100 \\
GLaS~\cite{glas}                                     & Histopathology & 165  & 72  & 13  & 80  \\
\textsuperscript{$\S$}PROMISE-12~\cite{promise}         & Prostate MRI & 1473 & 778 & 277 & 418 \\
\midrule
UDIAT~\cite{UDIAT}                                   & Breast US & 163  & – & – & 163 \\
Mendeley~\cite{mendeley}                             & Breast US & 250  & – & – & 250 \\
CVC-ColonDB~\cite{CVCColonDB}                        & Colonoscopy  & 380  & – & – & 380 \\
\bottomrule
\end{tabular}
\begin{tablenotes}
\footnotesize
\item[\textsuperscript{$\dagger$}]From the 780 images of BUSI, only the 647 lesion-containing samples are used.
\item[\textsuperscript{$\ddagger$}]Only the tooth-level annotations from the TDD are used.
\item[\textsuperscript{$\S$}]Only the extracted 2D axial slices containing the prostate region are considered from PROMISE-12.
\end{tablenotes}
\end{threeparttable}
\end{table}

\vspace{2mm}
\subsubsection{Evaluation Metrics}
To comprehensively and objectively assess the proposed architecture, Intersection over Union (IoU) has been employed as the primary evaluation metric. Additionally, Dice similarity index (DICE), pixel-wise accuracy (ACC), F1-score (F1), and average symmetric surface distance (ASSD) are reported.

\begin{table*}[t]
\centering
\scriptsize
\caption{Quantitative comparison of segmentation performance on BUSI and Tufts Dental Dataset (TDD). 
All values denote image-wise test-set means. Full metric results, including precision and recall, are provided in Supplementary Table~S1.
Best, second-best, and third-best scores are highlighted in \textcolor{red}{red}, \textcolor{blue}{blue}, and \textcolor{brown}{brown}, respectively.}
\label{tab:busi_tdd_metrics}
\setlength{\tabcolsep}{8.3pt}
\renewcommand{\arraystretch}{1.1}
\begin{tabular}{l|l|c|ccccc|ccccc}
\hline
\multirow{2}{*}{\textbf{\makecell{Domain/\\Strategy}}} 
& \multirow{2}{*}{\textbf{Model}} 
& \multirow{2}{*}{\textbf{Year}} 
& \multicolumn{5}{c|}{\textbf{BUSI}} 
& \multicolumn{5}{c}{\textbf{TDD}} \\ 
\cline{4-13}
& & 
& IoU$\uparrow$ & DICE$\uparrow$ & ACC$\uparrow$ & F1$\uparrow$ & ASSD$\downarrow$
& IoU$\uparrow$ & DICE$\uparrow$ & ACC$\uparrow$ & F1$\uparrow$ & ASSD$\downarrow$ \\ 
\hline \hline

\multirow{8}{*}{\makecell[l]{Spatial/\\CNN}}
& UNet~\cite{unet} & 2015 
& 0.6583 & 0.7618 & 0.9485 & 0.7958 & 4.0041 
& 0.7731 & 0.8736 & 0.9723 & 0.8794 & 1.0549 \\

& UNet++~\cite{unet++} & 2019 
& 0.7614 & 0.8426 & 0.9643 & 0.8662 & 2.6828 
& 0.8049 & 0.9215 & 0.9809 & 0.9120 & 0.8811 \\

& ResUNet++~\cite{resunet++} & 2019 
& 0.7881 & 0.8764 & 0.9688 & 0.8847 & 2.1816 
& 0.8239 & 0.8952 & 0.9790 & 0.9053 & 0.8040 \\

& CE-Net~\cite{cenet} & 2019 
& \textcolor{blue}{0.8217} & 0.8799 & \textcolor{blue}{0.9765} & \textcolor{blue}{0.9066} & \textcolor{brown}{1.8718} 
& 0.8335 & 0.9200 & 0.9803 & 0.9205 & 0.7656 \\

& CPFNet~\cite{cpfnet} & 2020  
& 0.8074 & 0.8738 & 0.9727 & 0.8720 & 2.0494 
& 0.8353 & \textcolor{brown}{0.9217} & 0.9811 & \textcolor{blue}{0.9222} & \textcolor{brown}{0.7602} \\

& PraNet~\cite{pranet} & 2020  
& 0.7819 & 0.8512 & 0.9704 & 0.8845 & 2.3526 
& 0.7547 & 0.8668 & 0.9654 & 0.8683 & 4.5301 \\

& AAU-Net~\cite{aaunet} & 2022  
& 0.7930 & 0.8570 & 0.9657 & 0.8838 & 2.2345 
& 0.8150 & 0.9169 & 0.9800 & 0.9176 & 0.7787 \\

& Twin-SegNet~\cite{twinsegnet} & 2024  
& \textcolor{brown}{0.8145} & \textcolor{blue}{0.8862} & \textcolor{brown}{0.9751} & 0.8917 & \textcolor{blue}{1.8523} 
& \textcolor{blue}{0.8487} & \textcolor{blue}{0.9293} & 0.9153 & 0.9184 & 1.4607 \\

\hline

\multirow{2}{*}{\makecell[l]{Spatial/\\Transformer}}
& TransAttUNet~\cite{transattunet} & 2024  
& 0.7627 & 0.8480 & 0.9628 & 0.8712 & 2.6033 
& \textcolor{brown}{0.8385} & 0.9177 & \textcolor{brown}{0.9816} & \textcolor{brown}{0.9215} & \textcolor{blue}{0.7463} \\

& Swin-Unet~\cite{swinunet} & 2022
& 0.7441 & 0.8310 & 0.9713 & 0.8319 & 2.0722
& 0.8135 & 0.8976 & 0.9760 & 0.8976 & 0.7768 \\

\hline

\multirow{3}{*}{\makecell[l]{Spectral/\\CNN}}
& SF-UNet~\cite{sfunet} & 2024
& 0.7717 & 0.8543 & 0.9703 & 0.8543 & 5.3772
& 0.8263 & 0.8992 & 0.9789 & 0.8992 & 1.2133 \\

& FDE-Net~\cite{fdenet} & 2026
& 0.7388 & 0.8243 & 0.9663 & 0.8243 & 6.4396
& 0.8160 & 0.8823 & 0.9810 & 0.8823 & 1.1360 \\

& MEW-UNet~\cite{mewunet} & 2022  
& 0.8059 & 0.8599 & 0.9742 & \textcolor{brown}{0.9016} & 1.9447 
& 0.8287 & 0.9163 & 0.9792 & 0.9168 & 1.0137 \\

\hline

\makecell[l]{Spectral/\\Transformer}
& \rule{0pt}{3.0ex}EW-ViT~\cite{ew-vit} & 2025
& 0.7543 & 0.8392 & 0.9701 & 0.8392 & 5.5361
& 0.7810 & 0.8608 & 0.9766 & 0.8608 & 1.3751 \\

\hline

\multirow{2}{*}{\makecell[l]{Dual/\\CNN}}
& DBLNet~\cite{dblnet} & 2024
& 0.8010 & \textcolor{brown}{0.8817} & 0.9721 & 0.8817 & 8.5061
& 0.8259 & 0.8907 & \textcolor{blue}{0.9818} & 0.8907 & 2.0461 \\

& Y-Net~\cite{ynet} & 2021  
& 0.7721 & 0.8603 & 0.9629 & 0.8752 & 2.3571 
& 0.7916 & 0.9162 & 0.9796 & 0.9169 & 0.8853 \\

\hline

\makecell[l]{Ours}
& \rule{0pt}{3ex}\textbf{Phi-SegNet} & 2026 
& \textcolor{red}{0.8454} & \textcolor{red}{0.9147} & \textcolor{red}{0.9806} & \textcolor{red}{0.9198} & \textcolor{red}{1.4999}
& \textcolor{red}{0.8537} & \textcolor{red}{0.9329} & \textcolor{red}{0.9831} & \textcolor{red}{0.9334} & \textcolor{red}{0.6773} \\

\hline
\end{tabular}
\end{table*}

\subsection{Comparison with SOTA Methods}
To assess the performance of the proposed Phi-SegNet, a comprehensive quantitative evaluation was conducted against sixteen state-of-the-art segmentation architectures, as summarized in Tables~\ref{tab:busi_tdd_metrics}--\ref{tab:promise_metrics}. The compared methods cover a diverse spectrum of segmentation strategies, including spatial CNN-based models such as UNet~\cite{unet}, UNet++~\cite{unet++}, ResUNet++~\cite{resunet++}, CE-Net~\cite{cenet}, CPFNet~\cite{cpfnet}, PraNet~\cite{pranet}, AAU-Net~\cite{aaunet}, and Twin-SegNet~\cite{twinsegnet}; spatial Transformer-based models such as TransAttUNet~\cite{transattunet} and Swin-Unet~\cite{swinunet}; spectral CNN-based models such as SF-UNet~\cite{sfunet}, FDE-Net~\cite{fdenet}, and MEW-UNet~\cite{mewunet}; the spectral Transformer-based EW-ViT~\cite{ew-vit}; and dual CNN-based methods including DBLNet~\cite{dblnet} and Y-Net~\cite{ynet}. To ensure fairness and consistency, all competing methods were reimplemented by adapting their official codebases (if available) and trained under a unified protocol across the five datasets. Complementing these quantitative evaluations, Fig.~\ref{fig:qualitative_result} presents qualitative comparisons of Phi-SegNet against five top-performing models, demonstrating its superior boundary adherence and structural fidelity.

\subsubsection{Evaluation on Breast Lesion Segmentation}
As shown in Table~\ref{tab:busi_tdd_metrics}, Phi-SegNet achieves state-of-the-art performance on the BUSI dataset, attaining the highest IoU and F1-score of 84.54\% and 91.98\%, respectively. Compared to the second-best performing model, CE-Net, our model yields consistent improvements across all metrics, with a gain of +2.37\% in IoU, +1.32\% in F1, while reducing the ASSD from 1.8718 to 1.4999.

\begin{figure*}[htbp]
    \centering
    \includegraphics[width = 0.88\textwidth]{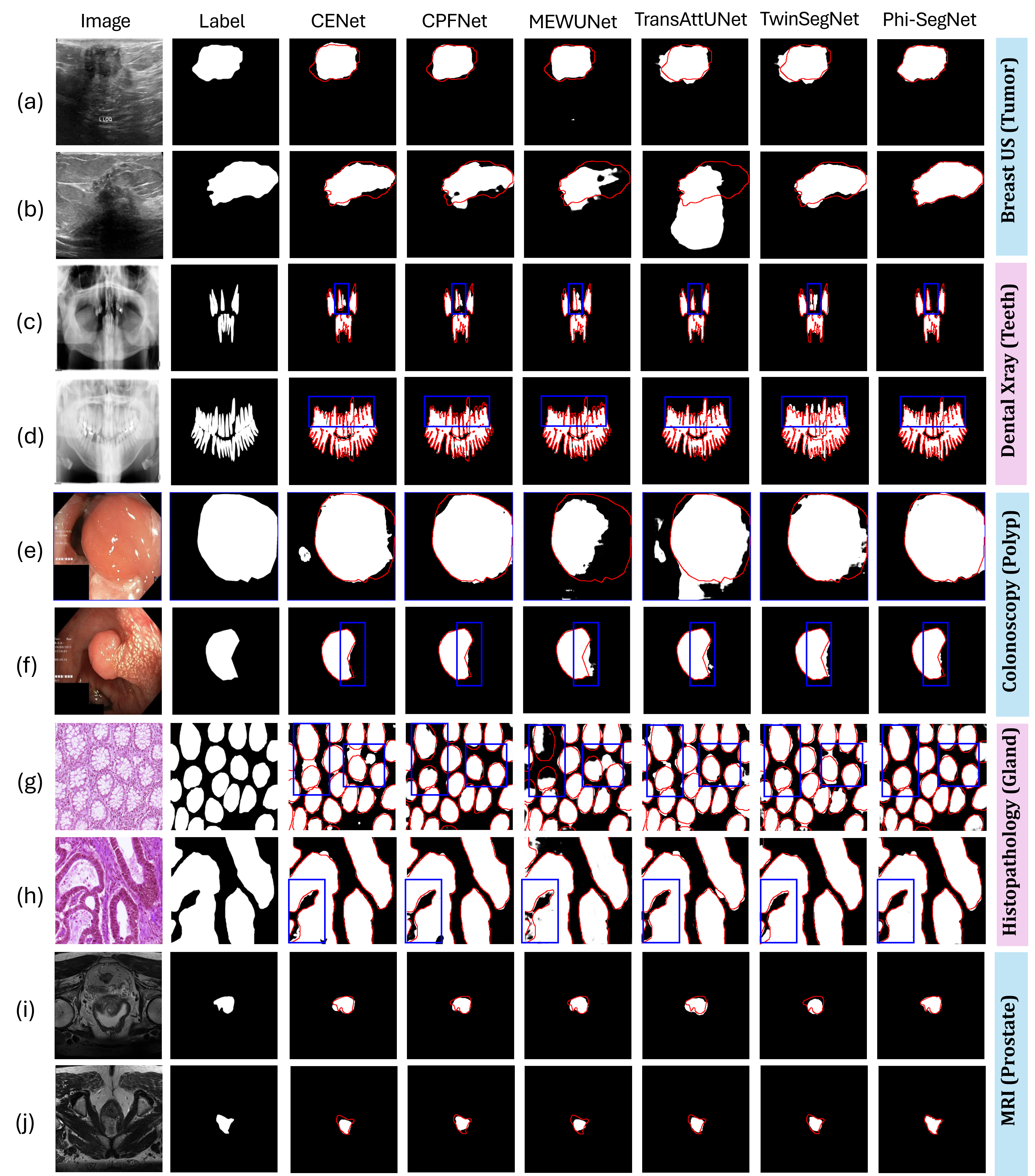}
    \caption{Visualization of segmentation results obtained by Phi-SegNet and other state-of-the-art architectures (two test cases from each dataset). Certain regions in some of the images have been marked using \textcolor{blue}{blue} bounding boxes to highlight the differences for the convenience of the reader.}
    \label{fig:qualitative_result}
\end{figure*}

From a visual standpoint, Fig.~\ref{fig:qualitative_result}(a) illustrates that several competing models struggle to accurately capture the complex and irregular outer contour of the malignant lesion. CENet, CPFNet, and MEWUNet undersegment the tumor, leaving large portions of the irregular boundary unaccounted for. TransAttUNet and Twin-SegNet over-segment the lesion by smoothing across concavities, ignoring localized boundary indentations. In contrast, Phi-SegNet achieves the best visual alignment with the ground truth, precisely tracing the jagged tumor edges. Notably, it is the only model to maintain contour continuity along the left and lower borders without leaking into adjacent tissues. Fig.~\ref{fig:qualitative_result}(b) further shows a challenging ultrasound case with weak lesion contrast and an irregular boundary. Several competing models either undersegment the lesion or produce spatially inconsistent masks, whereas Phi-SegNet accurately preserves the lesion extent and achieves the closest visual alignment with the ground-truth contour. These observations highlight Phi-SegNet's robustness in capturing heterogeneous lesion boundaries in breast ultrasound images.

\begin{table*}[t]
\centering
\scriptsize
\caption{Quantitative comparison of segmentation performance on Kvasir-SEG and GLaS datasets. 
All values denote image-wise test-set means. Full metric results, including precision and recall, are provided in Supplementary Table~S1.
Best, second-best, and third-best scores are highlighted in \textcolor{red}{red}, \textcolor{blue}{blue}, and \textcolor{brown}{brown}, respectively.}
\label{tab:kvasir_glas_metrics}
\setlength{\tabcolsep}{8.3pt}
\renewcommand{\arraystretch}{1.1}
\begin{tabular}{l|l|c|ccccc|ccccc}
\hline
\multirow{2}{*}{\textbf{\makecell{Domain/\\Strategy}}} 
& \multirow{2}{*}{\textbf{Model}} 
& \multirow{2}{*}{\textbf{Year}} 
& \multicolumn{5}{c|}{\textbf{Kvasir-SEG}} 
& \multicolumn{5}{c}{\textbf{GLaS}} \\ 
\cline{4-13}
& & 
& IoU$\uparrow$ & DICE$\uparrow$ & ACC$\uparrow$ & F1$\uparrow$ & ASSD$\downarrow$
& IoU$\uparrow$ & DICE$\uparrow$ & ACC$\uparrow$ & F1$\uparrow$ & ASSD$\downarrow$ \\ 
\hline \hline

\multirow{8}{*}{\makecell[l]{Spatial/\\CNN}}
& UNet~\cite{unet} & 2015 
& 0.6433 & 0.7566 & 0.9266 & 0.7976 & 5.3399 
& 0.5114 & 0.6598 & 0.5118 & 0.6767 & 68.4656 \\

& UNet++~\cite{unet++} & 2019 
& 0.7465 & 0.8344 & 0.9490 & 0.8642 & 3.1446 
& 0.7900 & 0.8543 & 0.8801 & 0.8897 & 3.3494 \\

& ResUNet++~\cite{resunet++} & 2019 
& 0.7140 & 0.8014 & 0.9399 & 0.8467 & 3.9315 
& 0.7109 & 0.8162 & 0.8297 & 0.8387 & 4.2321 \\

& CE-Net~\cite{cenet} & 2019 
& 0.8032 & 0.8701 & 0.9637 & \textcolor{brown}{0.8982} & 2.5477 
& 0.8253 & 0.8991 & 0.9002 & \textcolor{brown}{0.9057} & \textcolor{blue}{2.3485} \\

& CPFNet~\cite{cpfnet} & 2020  
& \textcolor{blue}{0.8162} & \textcolor{brown}{0.8794} & 0.9653 & \textcolor{blue}{0.9033} & 2.9053 
& \textcolor{blue}{0.8287} & \textcolor{blue}{0.9056} & \textcolor{blue}{0.9075} & \textcolor{blue}{0.9102} & 2.9475 \\

& PraNet~\cite{pranet} & 2020  
& 0.7550 & 0.8377 & 0.9471 & 0.8709 & 3.8235 
& 0.6726 & 0.8036 & 0.7791 & 0.8148 & 16.9096 \\

& AAU-Net~\cite{aaunet} & 2022  
& 0.7584 & 0.8351 & 0.9487 & 0.8644 & 3.4935 
& 0.7637 & 0.8627 & 0.8712 & 0.8765 & 3.3333 \\

& Twin-SegNet~\cite{twinsegnet} & 2024  
& 0.8109 & \textcolor{blue}{0.8826} & 0.9648 & 0.8934 & \textcolor{brown}{2.2324} 
& \textcolor{brown}{0.8254} & \textcolor{brown}{0.9006} & \textcolor{brown}{0.9007} & 0.9053 & \textcolor{brown}{2.8476} \\

\hline

\multirow{2}{*}{\makecell[l]{Spatial/\\Transformer}}
& TransAttUNet~\cite{transattunet} & 2024  
& 0.7746 & 0.8557 & \textcolor{brown}{0.9658} & 0.8840 & 2.6258 
& 0.8053 & 0.8866 & 0.8889 & 0.8961 & 3.2690 \\

& Swin-Unet~\cite{swinunet} & 2022
& \textcolor{brown}{0.8158} & 0.8790 & \textcolor{blue}{0.9712} & 0.8791 & \textcolor{blue}{1.9811}
& 0.7684 & 0.8650 & 0.8681 & 0.8649 & 3.1188 \\

\hline

\multirow{3}{*}{\makecell[l]{Spectral/\\CNN}}
& SF-UNet~\cite{sfunet} & 2024
& 0.7519 & 0.8321 & 0.9495 & 0.8321 & 9.1476
& 0.7163 & 0.8235 & 0.8266 & 0.8235 & 6.7045 \\

& FDE-Net~\cite{fdenet} & 2026
& 0.7740 & 0.8546 & 0.9578 & 0.8546 & 7.2013
& 0.7368 & 0.8401 & 0.8451 & 0.8401 & 5.8058 \\

& MEW-UNet~\cite{mewunet} & 2022  
& 0.7753 & 0.8540 & 0.9516 & 0.8835 & 3.1729 
& 0.7787 & 0.8685 & 0.8747 & 0.8792 & 3.3747 \\

\hline

\makecell[l]{Spectral/\\Transformer}
& \rule{0pt}{3.0ex}EW-ViT~\cite{ew-vit} & 2025
& 0.6568 & 0.7590 & 0.9268 & 0.7590 & 11.6677
& 0.7255 & 0.8343 & 0.8364 & 0.8343 & 6.0972 \\

\hline

\multirow{2}{*}{\makecell[l]{Dual/\\CNN}}
& DBLNet~\cite{dblnet} & 2024
& 0.7026 & 0.7943 & 0.9348 & 0.7943 & 21.2488
& 0.7610 & 0.8552 & 0.8636 & 0.8552 & 9.6896 \\

& Y-Net~\cite{ynet} & 2021  
& 0.7254 & 0.8170 & 0.9424 & 0.8501 & 3.9857 
& 0.7450 & 0.8478 & 0.8614 & 0.8688 & 3.6409 \\

\hline

\makecell[l]{Ours}
& \rule{0pt}{3ex}\textbf{Phi-SegNet} & 2026 
& \textcolor{red}{0.8496} & \textcolor{red}{0.9117} & \textcolor{red}{0.9761} & \textcolor{red}{0.9224} & \textcolor{red}{1.7547}
& \textcolor{red}{0.8383} & \textcolor{red}{0.9096} & \textcolor{red}{0.9079} & \textcolor{red}{0.9149} & \textcolor{red}{2.3477} \\

\hline
\end{tabular}
\end{table*}

\subsubsection{Evaluation on Teeth Segmentation}
Achieving the lowest ASSD score of 0.6733, Phi-SegNet demonstrates almost perfect boundary delineation. The quantitative representation in Table~\ref{tab:busi_tdd_metrics} further reveals that it achieves the best IoU of 85.37\% and F1-score of 93.34\% (+0.50\% and +1.50\% higher than the strongest competing model Twin-SegNet in terms of IoU and F1, respectively).

Qualitatively, Fig.~\ref{fig:qualitative_result}(c)--(d) present two challenging samples with crowded tooth arrangements and narrow inter-tooth spacing. Most models fail to capture sharper crown contours, especially around the cusp region in Fig.~\ref{fig:qualitative_result}(c). While Twin-SegNet produces a visually clean mask, it undersegments the region near the lower premolars and introduces false positives around the upper incisors. Similarly, in Fig.~\ref{fig:qualitative_result}(d), an oversegmentation artifact is clearly visible in the upper incisor region of Twin-SegNet’s output. Phi-SegNet, on the other hand, demonstrates the highest fidelity to the ground truth in both cases, maintaining intricate boundaries and sharp anatomical edges.

\subsubsection{Evaluation on Polyp Segmentation}
As reported in Table~\ref{tab:kvasir_glas_metrics}, Phi-SegNet achieves the best segmentation performance on the  Kvasir-SEG dataset, with an IoU of 84.96\%, F1-score of 92.24\%, and the lowest ASSD of 1.7547. Compared to the second-best model, CPFNet, which records 81.62\% IoU and 90.33\% F1-score, our model offers an improvement of +3.34\% in IoU and +1.91\% in F1, while reducing ASSD by 1.1506. 

The visual assessment in Fig.~\ref{fig:qualitative_result}(e)--(f) further corroborates these findings for the polyp segmentation task. Both samples present challenging cases with polyps occupying most of the field of view and showing minimal morphological separation from the surrounding mucosa. Specifically, the boundary between the distal margin of the polyp and the mucosal fold appears visually continuous, leading to MEW-UNet and Twin-SegNet undersegmenting along the right lateral borders whereas TransAttUNet introduces significant spatial leakage along the edges. Phi-SegNet, however, produces more restrained and anatomically plausible masks with superior localization and refined boundary adherence, effectively resolving structural ambiguities under low-contrast conditions.

\subsubsection{Evaluation on Gland Segmentation} 
As reported in Table~\ref{tab:kvasir_glas_metrics}, Phi-SegNet achieves the best overall performance on the GLaS dataset, with the highest IoU of 83.83\%, F1-score of 91.49\%, and the lowest ASSD of 2.3477. Compared with CPFNet, the strongest competing model, Phi-SegNet improves IoU by +0.96\%, Dice score by +0.40\%, and F1-score by +0.47\%.
In Fig.~\ref{fig:qualitative_result}(g)--(h), most models, including CE-Net, MEW-UNet, and Twin-SegNet, struggle to maintain boundary separation between closely located gland instances. These issues are especially prominent in the boxed regions (blue), where gland boundaries become blurred or collapse entirely. Although TransAttUNet and CPFNet perform relatively better, their boundary delineation remains inconsistent with noticeable pixel gaps. In contrast, Phi-SegNet resolves ambiguous gland boundaries more effectively.

\subsubsection{Evaluation on Prostate Segmentation}
On the PROMISE-12 dataset, Phi-SegNet achieves the best performance with an IoU of 83.62\%, F1-score of 91.50\%, and the lowest ASSD of 0.8926, outperforming CPFNet by +0.54\% IoU and +0.40\% F1 (see Table~\ref{tab:promise_metrics}). 

Visual assessment in Fig.~\ref{fig:qualitative_result}(i)--(j) reinforces these findings. In Fig.~\ref{fig:qualitative_result}(i), the prostate exhibits a compact, bean-shaped structure with a subtle indentation along the left margin, where weak boundary contrast causes most models to deviate. Twin-SegNet and CPFNet undersegment the lateral regions, whereas TransAttUNet and CE-Net slightly overshoot inferiorly.  Fig.~\ref{fig:qualitative_result}(j) presents another low-contrast MRI case where the prostate boundary is small and spatially subtle relative to the surrounding pelvic anatomy. Several competing models show contour mismatch and incomplete boundary adherence. Phi-SegNet, however, produces smooth and anatomically coherent contours with minimal leakage in both cases, preserving fine structural details and demonstrating robustness to low contrast and intensity inhomogeneity in pelvic MRI scans.

\begin{table}[t]
\centering
\scriptsize
\caption{Quantitative comparison of segmentation performance on the PROMISE-12 dataset. 
All values denote image-wise test-set means. 
Best, second-best, and third-best scores are highlighted in \textcolor{red}{red}, \textcolor{blue}{blue}, and \textcolor{brown}{brown}, respectively.}
\label{tab:promise_metrics}
\setlength{\tabcolsep}{2.2pt}
\renewcommand{\arraystretch}{1.05}
\begin{tabular}{l|l|c|ccccc}
\hline
\textbf{\makecell{Domain/\\Strategy}} 
& \textbf{Model} 
& \textbf{Year} 
& IoU$\uparrow$ & DICE$\uparrow$ & ACC$\uparrow$ & F1$\uparrow$ & ASSD$\downarrow$ \\ 
\hline \hline

\multirow{8}{*}{\makecell[l]{Spatial/\\CNN}}
& UNet~\cite{unet} & 2015 
& 0.6924 & 0.7976 & 0.9898 & 0.8247 & 1.8001 \\ 

& UNet++~\cite{unet++} & 2019 
& 0.8231 & 0.8966 & 0.9949 & 0.9031 & 0.9851 \\ 

& ResUNet++~\cite{resunet++} & 2019 
& 0.8186 & 0.8945 & 0.9949 & 0.9037 & 0.9932 \\ 

& CE-Net~\cite{cenet} & 2019 
& \textcolor{brown}{0.8293} & \textcolor{brown}{0.9015} & 0.9952 & 0.9096 & 0.9505 \\ 

& CPFNet~\cite{cpfnet} & 2020 
& \textcolor{blue}{0.8308} & \textcolor{blue}{0.9029} & \textcolor{brown}{0.9956} & \textcolor{brown}{0.9110} & \textcolor{blue}{0.9295} \\ 

& PraNet~\cite{pranet} & 2020 
& 0.8037 & 0.8867 & 0.9954 & 0.9037 & 0.9305 \\ 

& AAU-Net~\cite{aaunet} & 2022 
& 0.8263 & 0.8885 & 0.9953 & 0.8913 & 0.9573 \\ 

& Twin-SegNet~\cite{twinsegnet} & 2024 
& 0.8264 & 0.8972 & 0.9951 & 0.9062 & 0.9819 \\ 

\hline

\multirow{2}{*}{\makecell[l]{Spatial/\\Transformer}}
& TransAttUNet~\cite{transattunet} & 2024 
& 0.8256 & 0.8978 & \textcolor{blue}{0.9957} & \textcolor{blue}{0.9140} & \textcolor{brown}{0.9300} \\ 

& Swin-Unet~\cite{swinunet} & 2022
& 0.7785 & 0.8659 & 0.9934 & 0.8660 & 1.1266 \\

\hline

\multirow{3}{*}{\makecell[l]{Spectral/\\CNN}}
& SF-UNet~\cite{sfunet} & 2024
& 0.8068 & 0.8822 & 0.9942 & 0.8822 & 2.3597 \\

& FDE-Net~\cite{fdenet} & 2026
& 0.8042 & 0.8806 & 0.9942 & 0.8806 & 2.3664 \\

& MEW-UNet~\cite{mewunet} & 2022 
& 0.8186 & 0.8942 & 0.9953 & 0.9038 & 1.0170 \\ 

\hline
\makecell[l]{Spectral/\\Transformer}
& \rule{0pt}{3ex}EW-ViT~\cite{ew-vit} & 2025
& 0.7602 & 0.8532 & 0.9925 & 0.8532 & 2.9849 \\
\hline
\multirow{2}{*}{\makecell[l]{Dual/\\CNN}}
& DBLNet~\cite{dblnet} & 2024
& 0.8140 & 0.8876 & 0.9944 & 0.8876 & 4.0603 \\

& Y-Net~\cite{ynet} & 2021 
& 0.8203 & 0.8940 & 0.9950 & 0.9046 & 0.9959 \\ 

\hline

\makecell[l]{Ours}
& \rule{0pt}{3ex}\textbf{Phi-SegNet} & 2026 
& \textcolor{red}{0.8362} & \textcolor{red}{0.9055} & \textcolor{red}{0.9965} & \textcolor{red}{0.9150} & \textcolor{red}{0.8926} \\ 

\hline
\end{tabular}
\end{table}

\subsection{Evaluation of Generalization}
To assess the generalization capability of Phi-SegNet, we have trained a unified model (Phi-Seg\textsubscript{gen}) on a combined dataset encompassing the first five datasets listed in Table~\ref{tab:dataset_summary}. Its performance has been evaluated both on the merged test set and on each individual test set. 

\begin{table}[h]
\centering
\scriptsize
\caption{Performance of Phi-SegNet\textsubscript{gen}. Trained on the merged dataset, results reported for the merged test set as well as the individual test sets}
\label{tab:gen_metrics}
\setlength{\tabcolsep}{4.5pt}
\renewcommand{\arraystretch}{1.05}
\begin{tabular}{l|ccccccc}
\hline
\textbf{Dataset} & IoU$\uparrow$ & DICE$\uparrow$ & ACC$\uparrow$ & PRE$\uparrow$ & REC$\uparrow$ & F1$\uparrow$ & ASSD$\downarrow$ \\ 
\hline \hline
BUSI       & 0.7936 & 0.8681 & 0.9748 & 0.9061 & 0.8725 & 0.8681 & 0.2443 \\ 
Kvasir-SEG & 0.7835 & 0.8611 & 0.9576 & 0.8776 & 0.8962 & 0.8611 & 3.1044 \\ 
TDD        & 0.8324 & 0.9197 & 0.9800 & 0.9211 & 0.9194 & 0.9197 & 0.6909 \\ 
GLaS       & 0.6369 & 0.7704 & 0.8065 & 0.9209 & 0.6776 & 0.7704 & 6.2019 \\ 
PROMISE-12  & 0.8036 & 0.8841 & 0.9944 & 0.8910 & 0.9021 & 0.8841 & 1.1165 \\ 
\hline
\textbf{Merged} & 0.7897 & 0.8752 & 0.9674 & 0.8987 & 0.8809 & 0.8752 & 1.8546 \\ 
\hline
\end{tabular}
\end{table}

As summarized in Table~\ref{tab:gen_metrics}, Phi-Seg\textsubscript{gen} achieves a mean IoU of 78.97\% and Dice score of 87.52\% on the merged test set across modalities, indicating strong generalization without task-specific fine-tuning. Its performance closely follows that of dataset-specific trained models on TDD and PROMISE-12 and remains competitive on BUSI and Kvasir-SEG. Nevertheless, a noticeable drop is observed for GLaS (IoU = 63.69\%), likely due to its limited representation ($\approx$ 2.5\% of training data). 

\subsection{Unseen Cross-Dataset Evaluation}
To evaluate robustness beyond the training distribution, we have performed cross-dataset testing on three unseen datasets—UDIAT, Mendeley, and CVC-ColonDB. For each target domain, two variants are compared: the generalized model (Phi-Seg\textsubscript{gen}) trained on all modalities, and the modality-specific model (e.g., BUSI for ultrasound, Kvasir-SEG for colonoscopy). As shown in Table~\ref{tab:unseen_res}, modality-specific weights consistently yield superior segmentation across all datasets, confirming the perception that domain-relevant knowledge enhances performance. Despite having no prior exposure to the target datasets, \textbf{Phi-Seg}\textsubscript{BUSI} and \textbf{Phi-Seg}\textsubscript{Kvasir} maintain fair segmentation quality, indicating that the network effectively learns transferable structural priors such as lesion morphology and boundary continuity.

\begin{table}[h]
\centering
\scriptsize
\caption{Unseen cross-dataset evaluation of Phi-SegNet using generalized and modality-specific weights}
\label{tab:unseen_res}
\setlength{\tabcolsep}{2.7pt}
\renewcommand{\arraystretch}{1.05}
\begin{tabular}{l|l|ccccccc}
\hline
\textbf{Dataset} & \textbf{Variant} & IoU$\uparrow$ & DICE$\uparrow$ & ACC$\uparrow$ & PRE$\uparrow$ & REC$\uparrow$ & F1$\uparrow$ & ASSD$\downarrow$ \\ 
\hline \hline
\multirow{2}{*}{UDIAT} 
& Phi-Seg\textsubscript{gen} & 0.7043 & 0.7899 & 0.9823 & 0.7825 & 0.8764 & 0.7901 & 2.0798 \\ 
& Phi-Seg\textsubscript{BUSI} & 0.7796 & 0.8561 & 0.9876 & 0.8206 & 0.9168 & 0.8562 & 1.5046 \\ 
\hline
\multirow{2}{*}{Mendeley} 
& Phi-Seg\textsubscript{gen} & 0.1821 & 0.2114 & 0.8702 & 0.5740 & 0.2107 & 0.2115 & 12.7983 \\ 
& Phi-Seg\textsubscript{BUSI} & 0.6261 & 0.7261 & 0.9383 & 0.8490 & 0.7358 & 0.7261 & 5.2476 \\ 
\hline
\multirow{2}{*}{\shortstack{CVC\\Colon-DB}} 
& Phi-Seg\textsubscript{gen} & 0.5423 & 0.6214 & 0.9601 & 0.8039 & 0.6364 & 0.6215 & 3.9703 \\ 
& Phi-Seg\textsubscript{Kvasir}  & 0.5932 & 0.6622 & 0.9633 & 0.8341 & 0.6715 & 0.6622 & 3.5241 \\ 
\hline
\end{tabular}
\end{table}

\section{Discussion}
\subsection{Ablation Studies}
To evaluate the contribution of each architectural component, we conducted a series of ablation studies on the BUSI dataset. Starting from a baseline with an EfficientNet-B4 encoder–decoder and spatial loss ($\mathcal{L}_s$), we progressively introduced phase supervision loss ($\mathcal{L}_\varphi$), Bi-Feature Mask Former (BFMF) modules, inter-level attention blocks, and Reverse Fourier Attention (R$\mathcal{F}$A) modules to quantify their individual and combined effects. The corresponding model variants and results are summarized in Table~\ref{tab:ablation_busi}.

\subsubsection{Effect of Incorporating Phase Supervision Loss ($\mathcal{L}_\varphi$)}
Comparing the baseline (Row 1) with its phase-supervised counterpart (Row 2) in Table~\ref{tab:ablation_busi}, an increase in IoU from 0.7792 to 0.7905 and F1 from 0.8574 to 0.8672 can be observed, clearly indicating improved structural delineation. A similar absolute gain (+1.23\% IoU, +0.69\% F1) is observed when $\mathcal{L}_\varphi$ is applied on the BFMF-embedded baseline architecture (Row 3 vs. Row 4). These results confirm that phase supervision provides complementary boundary-aware guidance to spatial-domain learning.
\vspace{0.2 cm}
\begin{table}[t]
\centering
\caption{Ablation study on BUSI (Test Set) showing the effect of architectural modules and loss components}
\label{tab:ablation_busi}
\setlength{\tabcolsep}{4.3pt}
\renewcommand{\arraystretch}{1.01}
\begin{tabular}{l|c|c|c|c}
\toprule
\textbf{Model Variant} & \textbf{$\mathcal{L}_s$} & \textbf{$\mathcal{L}_\varphi$} & \textbf{IoU} & \textbf{F1} \\
\midrule
Baseline (EfficientNetB4) & \checkmark &  & 0.7792 & 0.8574 \\
Baseline (EfficientNetB4) & \checkmark & \checkmark & 0.7905 & 0.8672 \\
Baseline + BFMF & \checkmark &  & 0.8170 & 0.9037 \\
Baseline + BFMF  & \checkmark & \checkmark & 0.8293 & 0.9106 \\
Baseline + BFMF + Att & \checkmark & \checkmark & 0.8362 & 0.9150 \\
Baseline + BFMF + Att + R$\mathcal{F}$A ($\gamma$=3) & \checkmark & \checkmark & \textbf{0.8454} & \textbf{0.9198} \\
\bottomrule
\end{tabular}
\begin{tablenotes}
\footnotesize
\item[\textsuperscript{*}]*Whenever $\mathcal{L}_\varphi$ contributes to the total loss, $\varphi$-Conditioner is also present in the architecture to prepare the corresponding phase masks.
\end{tablenotes}
\end{table}

\subsubsection{Effect of Varying Loss Hyperparameters}
The loss-weight ratio between \(\mathcal{L}_{\varphi}\) and \(\mathcal{L}_s\) was empirically evaluated on the BUSI and Kvasir-SEG validation splits using the baseline model, as summarized in Table~\ref{tab:loss_busi}. These datasets were selected as representative validation cases from two different modalities, covering low-contrast breast lesion segmentation in ultrasound and polyp segmentation with variable colonoscopic appearance. \(\mathcal{L}_{\varphi}\) and \(\mathcal{L}_s\) guide the model toward complementary objectives. While \(\mathcal{L}_s\) guides region-level mask overlap and is treated as the primary segmentation objective, \(\mathcal{L}_{\varphi}\) acts as a secondary structural regularizer that encourages boundary and geometric alignment in the Fourier phase domain. Therefore, \(\beta\) was fixed at 1, and \(\alpha\) was varied using a coarse-to-fine empirical search. A relatively large phase-supervision hyperparameter \(\alpha\) can overemphasize phase consistency, causing the prediction to focus on structural alignment while weakening mask interior formation. In contrast, a very small \(\alpha\) makes the phase guidance ineffective. As shown in Table~\ref{tab:loss_busi}, \(\alpha=0.01\) with \(\beta=1\) provided the best overall balance in terms of IoU and F1. The same weights were used for all modalities in the main experiments to maintain a unified training protocol. In future work, adaptive loss-weighting strategies may be explored to automatically balance spatial and phase-domain supervision during training.

\begin{table}[h]
\centering
\caption{Effect of varying loss hyperparameters ($\alpha$, $\beta$) on segmentation performance using the baseline model}
\renewcommand{\arraystretch}{0.9}
\setlength{\tabcolsep}{9pt}
\begin{tabular}{cc|cc|cc}
\toprule
\multirow{2}{*}{$\alpha$} & \multirow{2}{*}{$\beta$} & 
\multicolumn{2}{c|}{\textbf{BUSI}} & 
\multicolumn{2}{c}{\textbf{Kvasir-SEG}} \\
\cmidrule(lr){3-4} \cmidrule(lr){5-6}
& & IoU & F1 & IoU & F1 \\
\midrule
1      & 1      & 0.0897 & 0.1852 & 0.1455 & 0.3479 \\
0.5    & 1      & 0.0766 & 0.1536 & 0.2394 & 0.4117 \\
0.1    & 1      & 0.7066 & 0.8274 & 0.7104 & 0.8485 \\
0.05   & 1      & 0.7158 & 0.8421 & 0.7463 & 0.8604 \\
0.025  & 1      & 0.7284 & 0.8426 & 0.7582 & 0.8736 \\
\textbf{0.01} & \textbf{1} & \textbf{0.7460} & \textbf{0.8520} & \textbf{0.7658} & \textbf{0.8772} \\
0.005  & 1      & 0.7258 & 0.8293 & 0.7505 & 0.8692 \\
\bottomrule
\end{tabular}
\label{tab:loss_busi}
\end{table}

\subsubsection{Effect of Incorporating BFMF Modules}
\begin{figure}[h]
    \centering
    \includegraphics[width=0.95\linewidth]{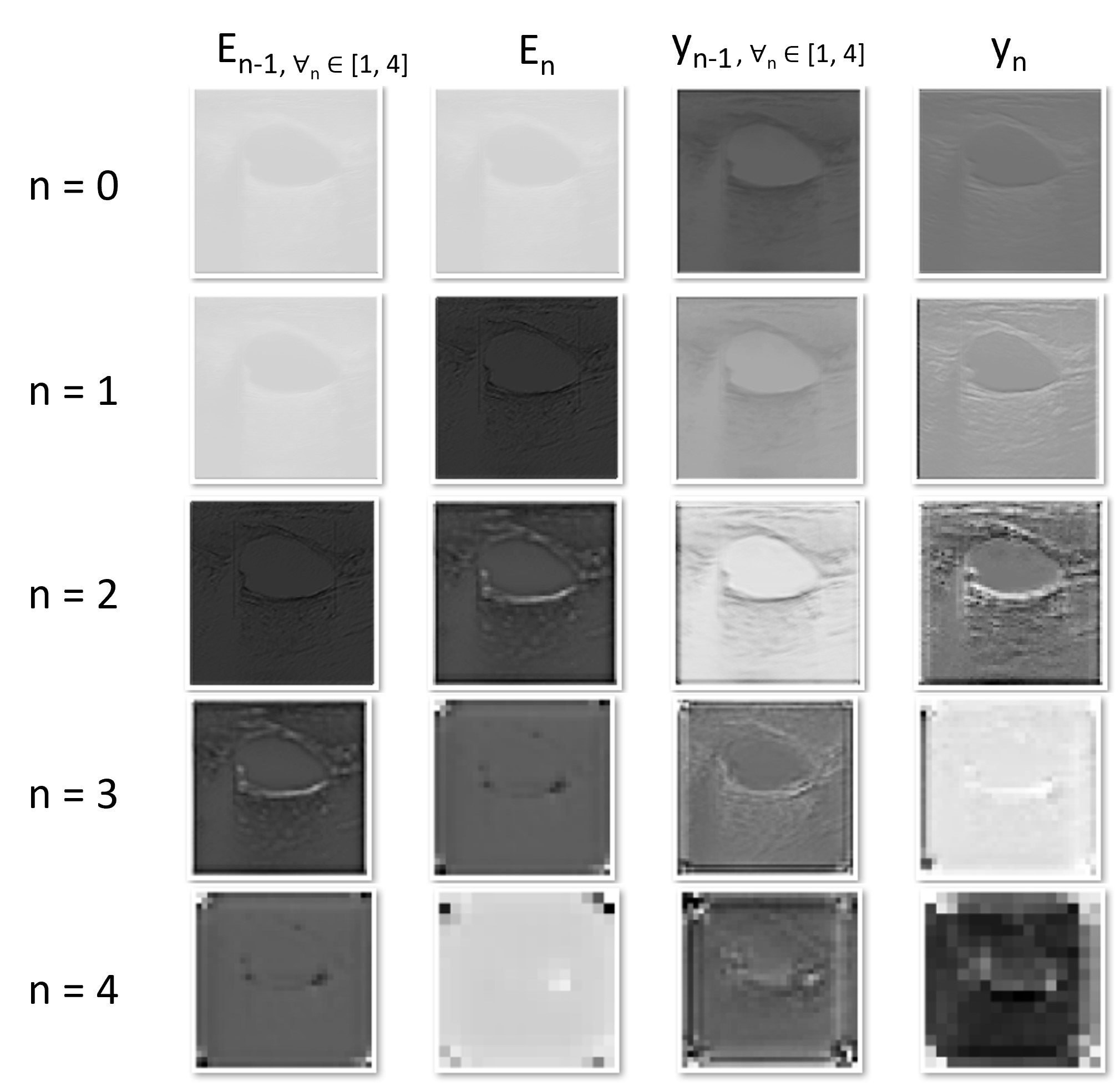}
    \caption{Visualization of BFMF feature maps (single-channel). From left to right: input features from $E_{n-1}$ (higher resolution), $E_n$ (lower resolution), and the corresponding BFMF outputs $y_{n-1}$ and $y_n$. Here, $n$ denotes the encoder level, with $n=0$ indicating the highest spatial resolution. All feature maps have been rescaled to $256 \times 256$ for better perception. Lesions appear more prominent in the fused outputs, demonstrating the effectiveness of BFMF.}

    \label{fig:bfmf_feat}
\end{figure}

As shown in Table~\ref{tab:ablation_busi}, integrating BFMF modules to the baseline significantly boosts performance, with IoU increasing from 0.7792 to 0.8170 and F1-score from 0.8574 to 0.9037. Under phase supervision (Row 2 vs. Row 4), a similar IoU gain (+3.88\%) is observed, confirming that BFMF effectively bridges the semantic gap between shallow and deep encoder features through multi-kernel contextual fusion. Fig.~\ref{fig:bfmf_feat} offers a visual justification of the behavior of BFMF. From the intermediate feature maps, it can be observed that compared to the encoders ($E_n$, $E_{n-1}$), the outputs of the BFMF ($y_n$ and $y_{n-1}$) depict enhanced contrast and better lesion focus, indicating successful modulation and feature refinement.

Although all experiments used \(256 \times 256\) input images for fair comparison, the proposed BFMF module is fully convolutional and is not restricted to this resolution. Since it operates on multi-scale encoder feature maps, BFMF can be applied to higher-resolution inputs without changing the overall network topology. For substantially larger images, the effective receptive field may be increased using a larger dilation rate or a larger kernel, such as replacing the current \(5\times5\) branch with \(7\times7\). This adjustment is optional and mainly useful when very fine structures or broader spatial dependencies must be captured, at the cost of higher computation.

\begin{figure*}[t]
    \centering
    \includegraphics[width=\textwidth]{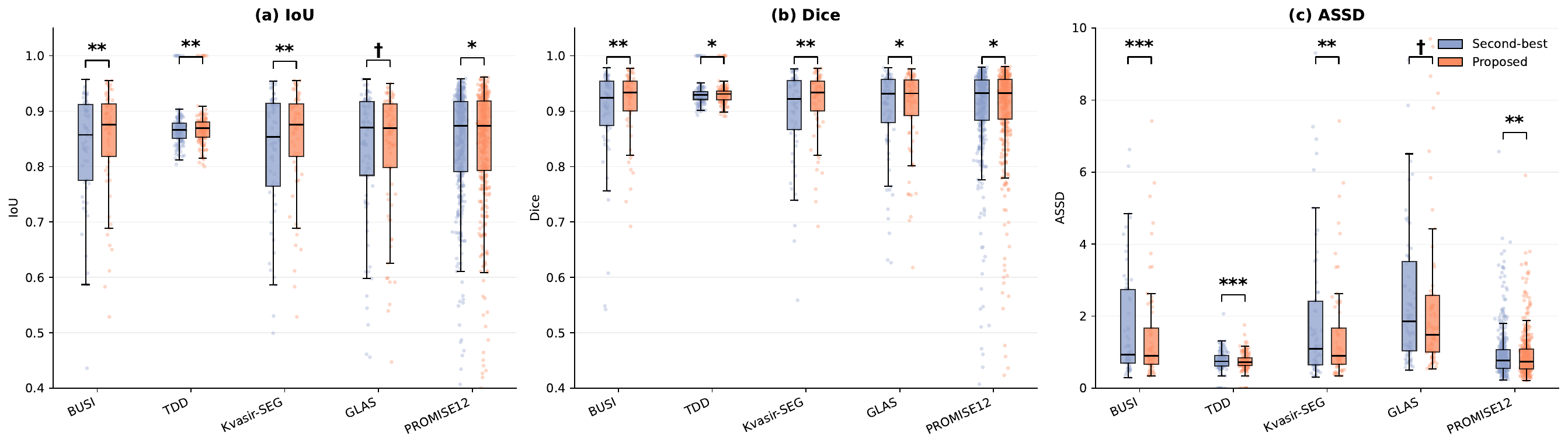}
    \caption{Image-wise statistical significance analysis comparing Phi-SegNet with the strongest competing method across five datasets. Boxplots show the distributions of (a) IoU, (b) Dice, and (c) ASSD. Statistical significance was assessed using a one-tailed paired Wilcoxon signed-rank test, with \(\dagger\), \(\star\), \(\star\star\), and \(\star\star\star\) denoting \(p<0.10\), \(p<0.05\), \(p<0.01\), and \(p<0.001\), respectively.}
    \label{fig:statistical_analysis}
\end{figure*}

\subsubsection{Effect of Incorporating R$\mathcal{F}$A Modules}
\begin{figure}[h]
    \centering
    \includegraphics[width=\linewidth]{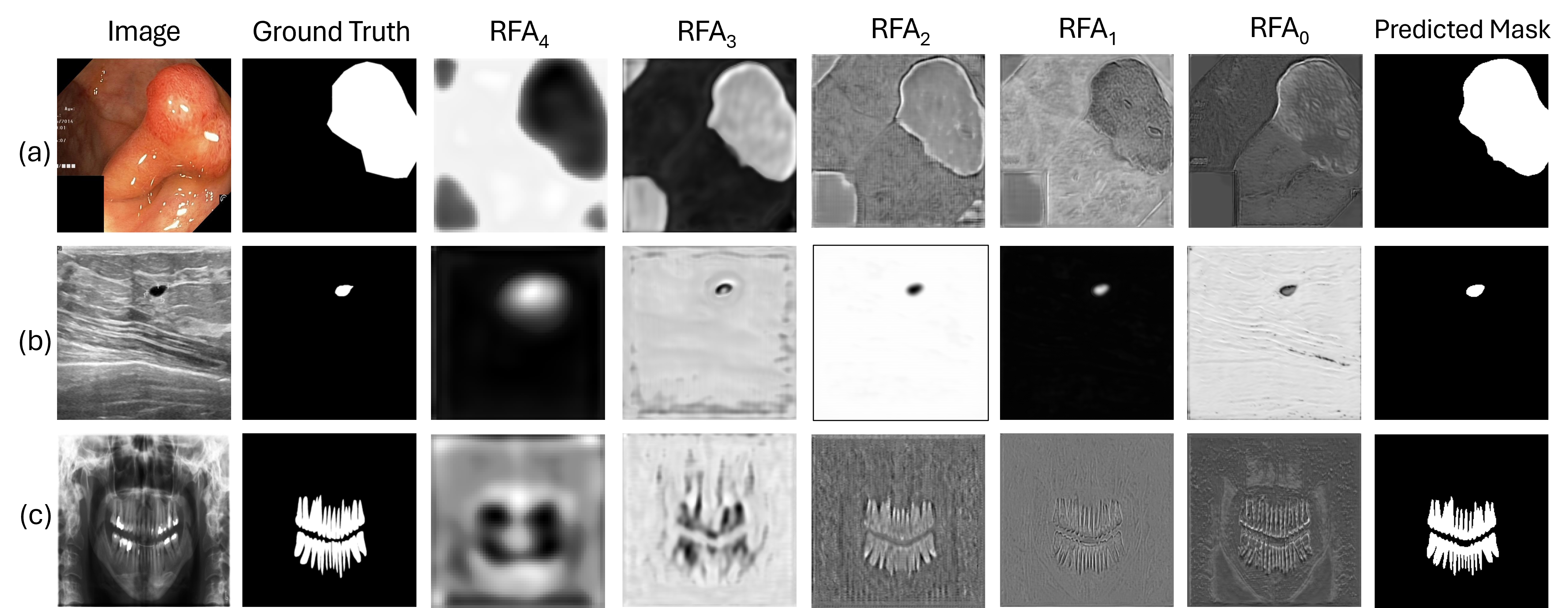}
    \caption{Feature maps from R$\mathcal{F}$A across decoder stages ($n = 4$ to $0$) in the final Phi-SegNet. As decoding progresses, spatial resolution increases and feature refinement improves significantly—culminating in sharper boundary localization at R$\mathcal{F}$A$_0$.}
    \label{fig:rfa_feature_maps}
\end{figure}

The Reverse Fourier Attention (R$\mathcal{F}$A) modules guide decoder features through frequency-domain modulation using a simple $3\times3$ low-pass filter (LPF). While high-pass filtering (HPF) might intuitively seem favorable for enhancing edge details, our experiments revealed the opposite (see Table~\ref{tab:rfa_gamma}). HPF degraded performance (IoU: 0.7442), likely amplifying high-frequency noise present in early decoder stages. To explore smoother spectral transitions, we also implemented a leaky $\gamma$-weighted LPF defined as:
\[
\Gamma(k,l)=\left(1+\sqrt{\frac{k^2+l^2}{M^2+N^2}}\right)^{-\gamma}, \quad \gamma>0,
\]
which gradually attenuates high-frequency components instead of imposing an abrupt cutoff. Although this formulation demonstrated marginal improvement (IoU: 0.7642), the standard LPF achieved the best performance (IoU: 0.7956, F1: 0.8662). This can be attributed to the effective suppression of high-frequency disturbances by the LPF, resulting in better structural clarity in downstream feature refinement. This hypothesis is visually supported in Fig.~\ref{fig:rfa_feature_maps}, which presents the saliency maps produced by R$\mathcal{F}$A ($\gamma = 3$) at multiple decoder depths. While R$\mathcal{F}$A$_4$ generates diffuse and noisy responses, progressively shallower modules (e.g., R$\mathcal{F}$A$_0$) exhibit sharper lesion boundaries and more localized attention. The phase-conditioned features can provide clinically meaningful interpretability by emphasizing boundaries of tumors, masses, lesions, and other anatomical inclusion regions. In many diagnostic tasks, boundary morphology is highly relevant, as irregular, poorly circumscribed, or infiltrative margins may support malignancy assessment and influence diagnosis, treatment planning, and prognosis. Therefore, the learned phase-conditioned maps can be viewed as boundary-emphasis representations that show whether the network attends to clinically important lesion margins. These maps may be further validated by radiologists through comparison with expert-annotated contours and clinically relevant margin characteristics.

Further analysis was conducted to assess the sensitivity of LPF-based RFA to different cutoff values, \(\gamma=\{3,5,9\}\). As shown in Table~\ref{tab:rfa_gamma}, \(\gamma=3\) appeared to be the optimal setting, achieving the most balanced trade-off between structural preservation and suppression of unstable residual responses on the BUSI validation set. Similar behavior was observed across the other evaluated datasets. Larger $\gamma$ values expanded the retained region but led to excessive smoothing, weakening critical boundary definitions. Overall, these findings underscore the importance of controlled frequency-domain modulation in phase-integrated learning pipelines, where both over-suppression and over-inclusion of spectral components can adversely affect segmentation fidelity. 

\begin{table}[t]
\centering
\caption{Impact of Fourier filter types and $\gamma$ parameters in R$\mathcal{F}$A. Results are reported on BUSI (Validation Set)}
\label{tab:rfa_gamma}
\begin{tabular}{l|c|c}
\toprule
\textbf{Model Variant} & \textbf{IoU} & \textbf{F1} \\
\midrule
Phi-SegNet Core + R$\mathcal{F}$A (No filter) & 0.7538 & 0.8380 \\ \hline
Phi-SegNet Core + R$\mathcal{F}$A (HPF: $\gamma=3$) & 0.7442 & 0.8235 \\
Phi-SegNet Core + R$\mathcal{F}$A (Leaky LPF: $\gamma=3$) & 0.7642 & 0.8474 \\
Phi-SegNet Core + R$\mathcal{F}$A (LPF: $\gamma=3$) & \textbf{0.7956} & \textbf{0.8662} \\
\hline
Phi-SegNet Core + R$\mathcal{F}$A (LPF: $\gamma=5$) & 0.7917 & 0.8639 \\
Phi-SegNet Core + R$\mathcal{F}$A (LPF: $\gamma=9$) & 0.7881 & 0.8609 \\
\bottomrule
\end{tabular}
\end{table}

\subsection{Statistical Significance Analysis}
To further assess whether the performance gains of Phi-SegNet were consistent at the image level, we conducted paired one-tailed Wilcoxon signed-rank tests between Phi-SegNet and the strongest competing method for each dataset. The comparator was selected as the second-best model based on mean IoU reported in Tables~\ref{tab:busi_tdd_metrics}--\ref{tab:promise_metrics}, resulting in CE-Net for BUSI, Twin-SegNet for TDD, and CPFNet for Kvasir-SEG, GLaS, and PROMISE-12. For every test image, IoU, Dice, and ASSD were computed for both Phi-SegNet and its matched comparator, and the paired metric distributions were compared. The resulting distributions and significance levels are visualized in Fig.~\ref{fig:statistical_analysis}.

At the conventional \(p<0.05\) significance threshold, Phi-SegNet showed significant improvements in IoU on BUSI, TDD, Kvasir-SEG, and PROMISE-12, significant Dice improvements across all five datasets, and significantly reduced ASSD on BUSI, TDD, Kvasir-SEG, and PROMISE-12. On GLaS, IoU and ASSD showed trend-level significance at \(p<0.10\). This weaker separation indicates a near-saturated setting in which both methods produce highly similar gland masks and boundary-distance distributions, consistent with Fig.~\ref{fig:qualitative_result}(g)--(h), where CPFNet already delineates the gland structures reasonably well and Phi-SegNet further improves localized boundary continuity in ambiguous regions. 

\subsection{Computational Complexity}
As reported in Table~\ref{tab:complexity}, Phi-SegNet contains 59.72M parameters and requires 81.91 GFLOPs per forward pass, with an average inference time of 20.41 ms. Although the parameter count exceeds that of most compared models, the increase arises from its joint spatial–frequency modeling and multi-scale aggregation in the BFMF modules. Compared to MEW-UNet and TransAttUNet, which rely on heavy backbones or full self-attention, Phi-SegNet achieves a notably better trade-off between accuracy and complexity. Architectures such as CPFNet and CE-Net are faster mainly due to custom lightweight encoders, whereas our use of EfficientNet-B4 backbone contributes to the higher FLOPs. Overall, Phi-SegNet maintains competitive runtime efficiency while delivering structurally robust segmentation, validating its practicality among state-of-the-art methods. 

\begin{figure}[h]
    \centering
    \includegraphics[width=.95\linewidth]{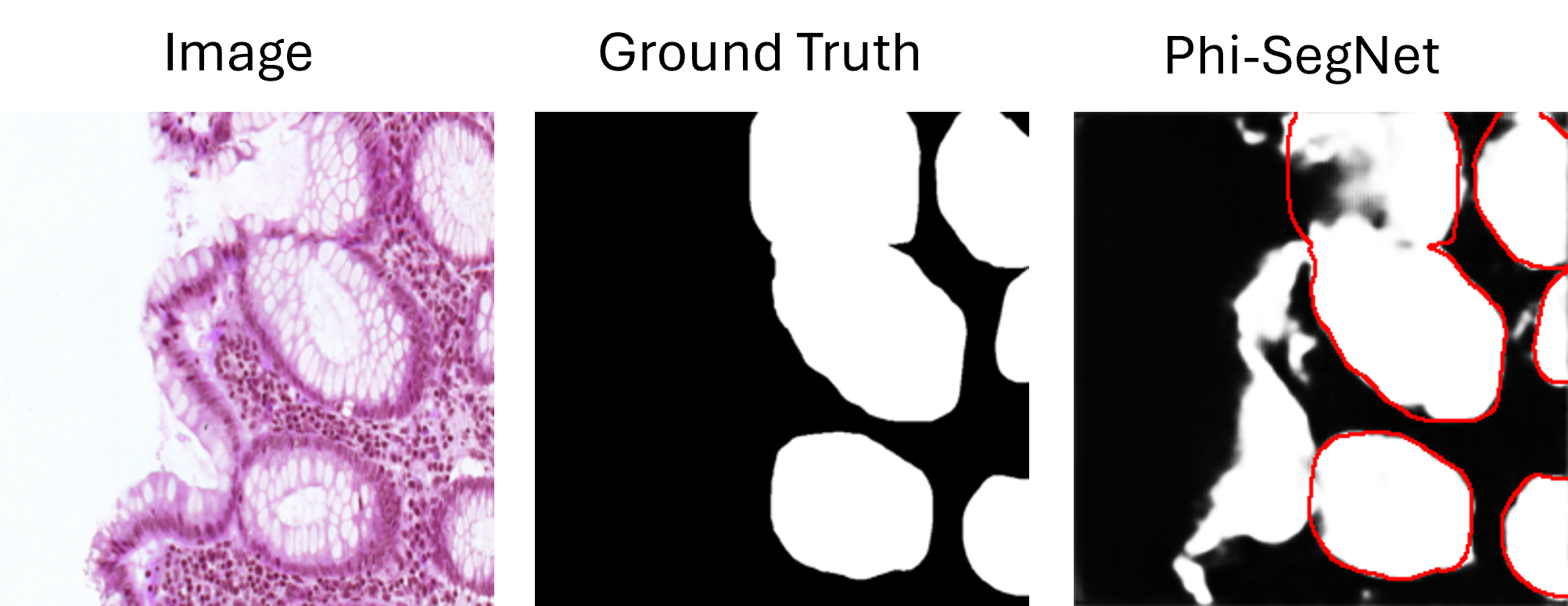}
    \caption{A representative failure case from the GLaS dataset.  The model incorrectly segments a bright non-tissue/background region as glandular structure, likely due to its visual similarity to glandular lumen and the presence of thin boundaries between adjacent glands.}
    \label{fig:failure_cases}
\end{figure}

\begin{table}[t]
\centering
\caption{Analysis of parameters, FLOPs, and inference time \\for different models}
\renewcommand{\arraystretch}{1.03}
\setlength{\tabcolsep}{4.5pt}
\begin{tabular}{l|c|c|c}
\hline
\textbf{Methods} & \textbf{\#Params (M)} & \textbf{FLOPs (G)} & \textbf{Inference Time (ms)} \\
\hline
\hline
UNet~\cite{unet} & 7.70 & 36.76 & 5.92 \\
UNet++~\cite{unet++} & 9.16 & 34.66 & 7.64 \\
ResUNet++~\cite{resunet++} & 4.06 & 15.75 & 7.07 \\
CE-Net~\cite{cenet} & 29.01 & 7.16 & 6.99 \\
CPFNet~\cite{cpfnet} & 43.27 & 8.36 & 9.63 \\
PraNet~\cite{pranet} & 32.55 & 6.92 & 18.81 \\
AAU-Net~\cite{aaunet} & 19.63 & 33.40 & 16.62 \\
Twin-SegNet~\cite{twinsegnet} & 71.78 & 16.32 & 22.57 \\
TransAttUNet~\cite{transattunet} & 25.97 & 88.75 & 13.82 \\
Swin-Unet~\cite{swinunet} & 27.16 & 17.73 & 9.17 \\
SF-UNet~\cite{sfunet} & 28.84 & 75.80 & 9.02 \\
FDE-Net~\cite{fdenet} & 8.83 & 49.86 & 11.72 \\
MEW-UNet~\cite{mewunet} & 140.27 & 41.23 & 95.68 \\
EW-ViT~\cite{ew-vit} & 35.32 & 7.01 & 32.90 \\
DBLNet~\cite{dblnet} & 20.95 & 41.95 & 7.34 \\
Y-Net~\cite{ynet} & 7.41 & 13.44 & 8.44 \\
\hline
\textbf{Phi-SegNet} & \textbf{59.72} & \textbf{81.91} & \textbf{20.41} \\
\hline
\end{tabular}
\label{tab:complexity}
\end{table}
\subsection{Failure Case Analysis}
Despite the overall robustness of Phi-SegNet, certain challenging cases remain difficult, particularly in cases with weak boundaries, partial occlusion, or underrepresented visual patterns. Fig.~\ref{fig:failure_cases} shows a representative failure case from GLaS. The model incorrectly identifies a bright non-tissue/background region as a glandular structure. This occurs because the background region and glandular lumen share similar visual characteristics: both appear as bright, low-texture areas adjacent to stained cellular structures. Moreover, the boundaries between neighboring glands are very thin, making gland-wise separation difficult and leading to partial merging of adjacent structures. This failure mode is consistent with the performance degradation observed for GLaS in the generalized setting, where histopathology samples constitute only a small fraction of the merged training data and provide limited morphology-specific variability. Increasing the input resolution or adopting patch-based training could improve the perception of fine gland boundaries and reduce such errors, but would require a different computational trade-off than the unified training setting used in this study.

\section{Limitations and Future Works}
While Phi-SegNet demonstrates compelling results, several limitations and opportunities remain. The EfficientNet-B4 backbone, though effective for hierarchical representation learning, increases computational and memory demands, constraining deployment in low-resource or real-time settings. Future work may investigate lightweight or compressed variants to improve efficiency. The fixed spectral filter in the R$\mathcal{F}$A modules ($\gamma=3$) may also limit adaptability across datasets with diverse frequency characteristics; adaptive spectral modulation parameterized by modality or anatomical context could enable data-driven frequency tuning. Furthermore, extending Phi-SegNet to weakly supervised learning and cross-domain adaptation may further enhance robustness and broaden clinical applicability across heterogeneous imaging environments.

\section{Conclusion}
This work presents Phi-SegNet, a CNN-based segmentation framework that integrates phase-aware supervision for enhanced structural delineation. The Bi-Feature Mask Former (BFMF) modules fuse adjacent encoder features to improve semantic coherence, while the Reverse Fourier Attention (R$\mathcal{F}$A) blocks refine decoder representations through phase-regularized features. Extensive evaluations across five imaging modalities confirm Phi-SegNet’s consistent state-of-the-art performance, particularly in fine-grained and low-contrast boundaries. Overall, Phi-SegNet reveals how phase-conditioned feature propagation and phase-based supervision can jointly drive more robust segmentation performance, paving the way for deeper integration of signal-level priors into deep learning frameworks for medical imaging.

\bibliographystyle{IEEEtran}
\bibliography{ref}
\end{document}